\documentclass[a4paper,fleqn]{cas-sc}

\usepackage[numbers]{natbib}

\newcommand{\AIGC}{\texttt{AIGC}}
\newcommand{\lora}{\texttt{LORA}}
\newcommand{\llava}{\texttt{LLAVA}}
\newcommand{\llm }{\texttt{LLM}}
\newcommand{\sdm}{\texttt{SDM}}
\newcommand{\sdxl}{\texttt{SDXL}}
\newcommand{\bsdxl}{\texttt{BSDXL}}

\newcommand{\asr}{\texttt{ASR}}
\newcommand{\nlp}{\texttt{NLP}}
\newcommand{\RL}{\texttt{RL}}
\newcommand{\rag}{\texttt{RAG}}
\newcommand{\cgpt}{\texttt{ChatGPT}}
\newcommand{\gpt}{\texttt{GPT}}
\newcommand{\duck}{\texttt{DuckDuckGo}}

\newcommand{\cbot}{\texttt{chatbot}}
\newcommand{\openai}{\texttt{OpenAI}}
\newcommand{\llama}{\texttt{LLama}}
\newcommand{\DNLM}{\texttt{DNLM}}

\newcommand{\ai}{\texttt{AI}}
\newcommand{\bioeng}{\texttt{Bio-Eng-LMM}}
\newcommand{\benoptex}{\texttt{BENOPTex}}
\newcommand{\nvidia}{\texttt{NVIDIA}}
\newcommand{\intel}{\texttt{Intel}}

\usepackage{graphicx}
\usepackage{svg}
\usepackage{hyperref}
\usepackage{cleveref}

\usepackage{amsmath,amsfonts}
\usepackage{natbib}
\usepackage{algorithmic}
\usepackage{algorithm}
\usepackage{array}
\usepackage[caption=false,font=normalsize,labelfont=sf,textfont=sf]{subfig}
\usepackage{textcomp}
\usepackage{stfloats}
\usepackage{url}
\usepackage{verbatim}
\usepackage{graphicx}
\usepackage{soul}

\begin{document}
\let\WriteBookmarks\relax
\def\floatpagepagefraction{1}
\def\textpagefraction{.001}
\shorttitle{\bioeng~\ai~Assist \cbot}
\shortauthors{Ali Forootani et~al.}

\title [mode = title]{
\bioeng~\ai~Assist: A Modular \cbot~Platform for Interdisciplinary Research and Education
}                      

\author[UFZ]{Ali Forootani*}[ orcid=0000-0001-7612-4016]
\ead{ali.forootani@ufz.de/aliforootani@ieee.org}
\author[UFZ]{Danial Esmaeili~Aliabadi}[orcid=0000-0003-2922-2400]
\ead{danial.esmaeili@ufz.de}
\author[UFZ,DBFZ,UNI]{Daniela Thr\"an}[orcid=0000-0002-6573-6401]
\ead{daniela.thraen@ufz.de}
\address[UFZ]{Helmholtz Centre for Environmental Research - UFZ, Permoserstraße 15, 04318 Leipzig, Germany}
\address[DBFZ]{DBFZ Deutsches Biomasseforschungszentrum gGmbH, Torgauer Strasse 116, 04347 Leipzig, Germany}
\address[UNI]{University Leipzig, Institute for Infrastructure and Resources Management, Grimmaische Str. 12, 04109 Leipzig, Germany}
\cortext[cor1]{Corresponding author}

\begin{abstract}
This article presents \bioeng~\ai~\cbot~Assist, a versatile \cbot~platform designed to support interactive learning and research across multiple disciplines. Originally developed for biomass research, the system’s capabilities extend to broader educational and scientific domains. It combines large language models (\llm s) with advanced tools for document analysis, real-time file and web data integration, image understanding, and speech recognition. Central to the platform is a Retrieval Augmented Generation (\rag) framework that enhances the accuracy and contextuality of responses by incorporating relevant external information. \bioeng~also supports image generation using diffusion models and secure web-based search and summarization. With a user-friendly interface and multimodal functionality (text, image, and voice), the platform offers dynamic assistance in academic settings. By simplifying access to complex data and enabling cross-disciplinary collaboration, \bioeng~helps users develop \ai~literacy while facilitating knowledge discovery and communication.
	
\end{abstract}

\begin{highlights}
\item Enhances personalized learning with AI-driven assistance.
\item Integrates real-time data for enriched research and analysis.
\item Fosters creativity through image generation and visualization tools.
\end{highlights}


\maketitle

\section{Introduction}
\label{sec:intro}
Recent advancements in artificial intelligence have propelled language models to unprecedented capabilities, exemplified by the emergence of multimodal large language models (\llm s) \citep{brown2020language, jose2024advancing}. These models, unlike their unimodal predecessors, process and understand multiple data modalities—including text, images, audio, and video—simultaneously \citep{wu2023multimodal}. This integration significantly enhances contextual understanding and response generation, surpassing the limitations of traditional models \citep{bai2024survey}. As \llm-guided \cbot s will become indistinguishable from human participants, they will likely play a prominent role as digital twins in social science research \citep{grossmann2023ai,bail2024can}.

Concurrently, Retrieval-Augmented Generation (\rag) techniques \citep{lewis2020retrieval} have emerged as a powerful paradigm for enhancing language models. By incorporating external knowledge sources, \rag~empowers models to access and leverage relevant information beyond their training data. When integrated into systems, \rag~facilitates the generation of more informative and accurate responses, bridging the gap between open-domain conversational agents and information retrieval systems \citep{zhao2024retrieval,xu2024enhancing}.

This intersection of multimodal \llm s and \rag~presents a promising avenue for the development of sophisticated conversational agents capable of handling complex queries and providing comprehensive responses. By combining the strengths of both approaches, future research can explore novel applications and address the challenges inherent in multimodal information processing and knowledge management.

\rag~ significantly enhances language model capabilities by combining generative and retrieval components. \rag~frameworks augment model outputs by incorporating relevant information retrieved from external sources. This hybrid approach overcomes limitations of static, pre-trained models by providing dynamic access to a broader knowledge base, resulting in more accurate and contextually relevant responses \citep{yu2024evaluation}.

The integration of multimodal \llm s and \rag~techniques into chatbot systems represents a substantial advancement in conversational \ai~\citep{panda2023exploring, oh2024language}.
 Traditional \cbot, often constrained by predefined responses and simple pattern matching, are evolving into sophisticated agents capable of rich, contextually aware interactions. By leveraging multimodal capabilities, these \cbot s can process and respond to complex queries involving text, images, and voice commands \citep{wang2024potential}. Simultaneously, \rag~empowers \cbot s with access to up-to-date information, enhancing response relevance and utility in real-world scenarios.

\subsection{Literature review}

Recent strides in machine learning are propelled by innovative algorithms, the rapid expansion of foundation models,
 and the availability of vast, high-quality datasets. Sequence-to-sequence tasks have witnessed a paradigm shift from Long Short-Term Memory (\texttt{LSTM}) networks \citep{hochreiter1997long} to Transformer-based models \citep{vaswani2017attention}. Similarly, image generation has evolved from Generative Adversarial Networks (\texttt{GANs}) \citep{goodfellow2014generative} to Latent Diffusion Models (\texttt{LDMs}) \citep{rombach2022high}. Foundation model architectures have expanded dramatically, with parameter counts soaring from millions to billions or even trillions. This rapid evolution is facilitated by rich, high-quality datasets that provide ample training data for effective model optimization.

 As a direct descendent of Deep Neural Language Models (\DNLM s), \llm s consume a massive amount of data, and their results are polished with the Reinforcement Learning (\RL) algorithms from
Human Feedback \citep{kucharavy2024deep}. \llm s,  which are the foundation for many famous products such as \openai's~\cgpt, leverage advanced machine learning techniques to generate human-like text, making them indispensable tools for various natural language processing (\nlp) applications \citep{chen2021evaluating, achiam2023gpt}.

Building upon the limitations of traditional \nlp~models,
Artificial Intelligence Generated Content (\AIGC) offers a paradigm shift through its ability to learn complex patterns directly from data \citep{cao2023comprehensive,liang2022foundations}. By leveraging massive datasets and advanced neural architectures, \AIGC~has demonstrated remarkable performance in various \nlp~tasks, including:
\begin{itemize}
    \item \textbf{Text and Code Generation:} Models such as \gpt~ (Generative Pre-trained Transformer) \citep{masalkhi2024google} and its successors, along with open-source alternatives like \llama~ \citep{touvron2023llama}, have demonstrated exceptional abilities in generating human-quality text and code.
    \item \textbf{Image Generation:} Models like \texttt{DALL-E} \citep{marcus2022very} and Stable Diffusion \citep{stable_diff} have revolutionized image creation, enabling the generation of highly realistic and diverse images from textual descriptions.
    \item \textbf{Video Production:} With models like \texttt{Sora} \citep{adetayo2024text} and \texttt{VideoGPT} \citep{yan2021videogpt}, AIGC has expanded its capabilities to video generation, producing coherent and visually appealing video content.
    \item \textbf{Image Understanding:} Models such as \llava~ \citep{liu2023improvedllava,liu2024llavanext, liu2023llava}, \texttt{MiniGPT} \citep{chen2023minigptv2, zhu2023minigpt}, and \texttt{MPlugOwl} \citep{ye2023mplug} have demonstrated impressive performance in understanding and interpreting visual content.
    \item \textbf{Automatic Speech Recognition:} Models like \texttt{Whisper} \citep{radford2023robust} and \texttt{Google}'s speech-to-text have significantly advanced the accuracy and efficiency of converting spoken language into text. 
\end{itemize}

\AIGC~tools employ advanced generative models, differentiating them from traditional content creation methods. These tools excel in producing high-quality, diverse outputs based on user-defined parameters \citep{wu2023ai}, offering unparalleled scalability. The term ``AIGC" underscores the AI-driven nature of content generation, signifying a paradigm shift from human-crafted or rule-based content to machine-produced materials \citep{xing2024ai}. The implications of \AIGC~are profound, spanning industries from entertainment and marketing to software development and scientific research \citep{lei2000industry}.

\begin{figure*}[t]
\centering
\includegraphics[width=0.8\linewidth]{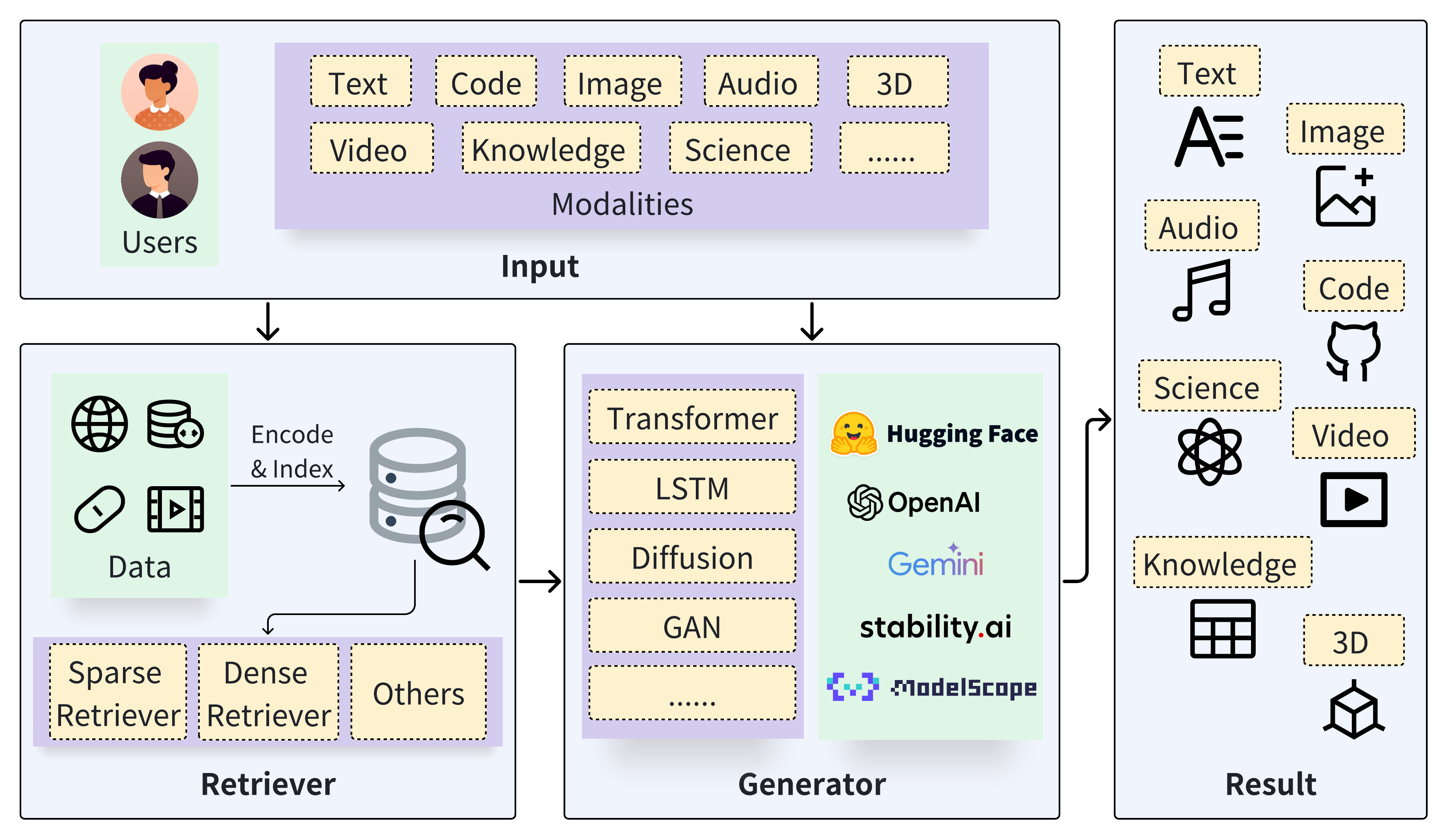}
\caption{A typical \rag~architecture functions as follows: The user submits queries in various formats, which are fed into both the retriever and the generator. The retriever extracts relevant information from data sources, and the generator uses this information to produce outputs in different formats.\citep{zhao2024retrieval}.}
\label{fig:ragoverview}
\end{figure*}

A typical \rag~process, illustrated in Figure \cref{fig:ragoverview}, involves a retriever fetching relevant information based on a given query, narrowing down the response space, as \llm s are trained on vast volumes of data, much of which is irrelevant to the tasks being requested. This retrieved data is subsequently integrated into the generator to enhance output quality. Integration methods encompass several strategies: (i) direct incorporation as additional input \citep{guu2020retrieval, lewis2020retrieval}; (ii) integration as latent representations during generation \citep{izacard2020leveraging, borgeaud2022improving}; (iii) contribution as logits\footnote{The raw, unnormalized output values generated by classification models in neural network systems.} to the final output, i.e. information is transformed into a format compatible with the model's output layer \citep{khandelwal2019generalization, he2021efficient}; (iv) guiding or skipping generation steps \citep{he2023rest}. By manipulating logits and generation steps, models can influence the final output \citep{khandelwal2019generalization, he2021efficient, he2023rest, baghdasaryan2024knowledge}. Building upon the foundational \rag~approach, researchers are continuously refining both the individual components of the system, such as the methods for retrieving and incorporating information, and the overall process to optimize performance and effectiveness. Individual components of a \rag~system typically include a retriever, which identifies relevant information, and a generator, which produces the final output.

Originally designed for text generation \citep{lewis2020retrieval, najafi2024turkishbertweet},
the \rag~framework has demonstrated adaptability across diverse data modalities, encompassing code, audio, images, video, 3D models, knowledge bases, and scientific data \citep{parvez2021retrieval, ahmad2021unified, zhou2022docprompting, koizumi2020audio, huang2023make, tseng2020retrievegan, sarto2022retrieval, ramos2023smallcap, chen2023retrieval, xu2024retrieval, seo2024retrieval, zhang2023remodiffuse, hu2022logical, DBLP:conf/emnlp/HuangKZ21, DBLP:conf/emnlp/DasZTGPLTPM21, wang2022retrieval, jin2023genegpt}. While the core \rag~process remains consistent, domain-specific adaptations, particularly in data augmentation and model selection, are essential for optimal performance.

Multimodal \cbot s incorporating \rag~offer transformative potential,
particularly in education and research. In the realm of education, these \cbot~create immersive learning experiences by processing diverse input formats (text, images, audio, video), fueling creative problem-solving performance in students \citep{urban2024chatgpt}. They provide personalized support tailored to individual learning styles and enhance accessibility for students with disabilities through features such as text-to-speech and visual explanations. Within research, multimodal \rag~\cbot~push boundaries by exploring novel integrations of diverse data. They contribute to identifying research challenges, establishing evaluation benchmarks, and fostering interdisciplinary collaboration. Ultimately, these advancements lay the groundwork for practical applications benefiting both educators and learners \citep{darvishi2024impact}.

Developing a multimodal large language model (\llm)-based \ai~assistant chatbot for educational purposes necessitates the integration of diverse data formats, including text, images, audio, and potentially video. Essential components for such a system encompass:

\begin{itemize}
    \item \textbf{Language Model:} A foundation \llm, such as \gpt~or a similar model, capable of understanding and generating human-like text.
    \item \textbf{Multimodal Capabilities:} The ability to process and generate multiple data formats, including text, images, audio, and potentially video.
    \item \textbf{User Interface:} An interactive \cbot~interface designed for both students and educators.
    \item \textbf{Backend Infrastructure:} Robust server, database, and API support for the \cbot s operation.
    \item \textbf{Educational Content:} A curated collection of content across various subjects.
\end{itemize}

These components are essential parts of any engaging story, which is continuously being used in educational environments \citep{wu2020systematic,chou2015prezi}. It has also been reported that converting complex topics to stories (using a whiteboard or a canvas) can assist even primary school students with no background knowledge about energy systems to develop a visual record of concepts \citep{Chappin}.

\subsection{Motivation and contribution}

The creative potential of \llm s extends to various educational projects, from artistic endeavors to technical experiments. Students can leverage the model's capabilities to generate novel visual content, thereby stimulating creativity and exploration within artistic and design disciplines. Furthermore, the technical challenges associated with fine-tuning and optimizing diffusion models provide valuable opportunities for students to develop problem-solving skills and proficiency in programming and data processing. These experiences are instrumental in preparing students for careers in \ai~and related fields. For example, the versatility of \sdm~enables their application across diverse academic disciplines. In historical studies, for instance, students can generate reconstructions of ancient artifacts or historical events, while in literature, they can visualize scenes or characters described in texts. This cross-disciplinary utility fosters collaborative learning environments where students from different fields can work together on integrative projects, thus promoting a holistic educational experience.

\bioeng~addresses key gaps in bio-mass research by offering a versatile, multimodal \ai~platform that enhances both data access and analytical capabilities. Through its \rag~functionality, the system allows researchers to seamlessly retrieve and process diverse sources of information, including preprocessed documents, user-uploaded files, and real-time web data. This broadens the scope of research inputs, streamlining data collection and analysis. Additionally, the platform's integration of image generation via \sdm~and image understanding through \llava~enables researchers to visualize complex bio-mass processes, such as decomposition or metabolic pathways, improving interpretative analysis. \bioeng~also facilitates cross-disciplinary collaboration, allowing researchers from various fields to engage with bio-mass studies through integrative projects. Furthermore, it promotes technical skill development in \ai, encouraging researchers to fine-tune models and apply advanced data manipulation techniques, thereby fostering innovation and problem-solving within bio-mass research.

This article introduces \bioeng, a developed multimodal \cbot~equipped with the following core functionalities: 

\begin{itemize}
    \item \textbf{\ai~Assistant:} Offers basic conversational capabilities akin to \gpt.
    \item \textbf{Retrieval Augmented Generation (\rag):} Accesses information from diverse sources, including preprocessed documents, user-uploaded files, and real-time web data, to enhance response quality.
    \item \textbf{Image Processing:} Leverages Stable Diffusion by Stability (\sdm) \ai~for image generation and \llava~for image content understanding and response generation.
    \item \textbf{Internet Search and Summarization:} Employs \duck~for search results and provides summaries of website content or documents. 
\end{itemize}

\section{Background}

This section provides an overview of the core tools underpinning the \bioeng~platform. We dive into the \rag ~architecture, examining both generators and retrievers commonly employed in contemporary \ai-generated content (AIGC). Additionally, we explore the integration of \llava~for image comprehension, Stable Diffusion, specifically Low Rank Adaptation (\lora) and Stable Diffusion Lightning (\sdxl), for image generation, and \texttt{Whisper-Base.en}, which is a transformer-based model, for robust speech recognition.

\subsection{Overview on \rag~and its foundation}

The \rag~system comprises two primary components: a retriever and a generator, as illustrated in \cref{fig:ragoverview}. The system operates through a three-step process:

\begin{itemize}
    \item Query Processing: The retriever processes an input query and identifies relevant information within a designated data repository.

    \item Information Augmentation: The system integrates the original query with the retrieved information using a specified augmentation technique.

    \item Content Generation: The generator leverages the augmented input to produce the final output. 
    
\end{itemize}

The retriever's role is to locate and extract relevant data, while the generator's task is to create the requested content based on the augmented information.

\subsubsection{Generator}

At the heart of the \rag~system lies the crucial generation module. This component employs diverse generative models tailored to specific tasks including (i) Text-to-text tasks such as \llama; (ii) Image-to-text tasks such as \texttt{VisualGPT} \citep{DBLP:conf/cvpr/ChenGY0E22}; (iii) Text-to-image tasks such as Stable Diffusion \citep{DBLP:conf/cvpr/RombachBLEO22}; (iv) Text-to-code tasks such as \texttt{Codex} \citep{DBLP:journals/corr/abs-2107-03374}. 

In the context of \rag~systems, four generator types are particularly prevalent: (i) Transformer models; (ii) Long Short-Term Memory (\texttt{LSTM}) networks; (iii) Diffusion models; (iv) Generative Adversarial Networks (\texttt{GAN}s). Each of these generators brings unique strengths to the table, enabling \rag~systems to tackle a wide array of content generation challenges across different modalities and applications.

\begin{figure*}[t]
\centering
\includegraphics[width=0.5\linewidth]{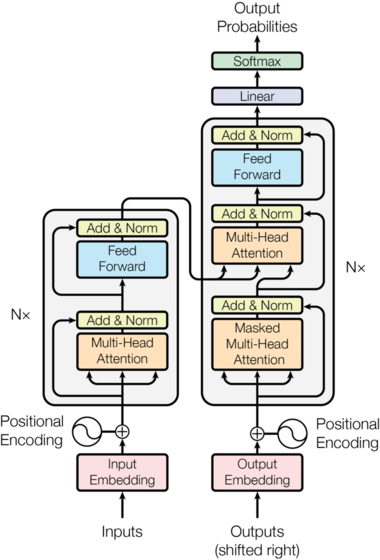}
\caption{Transformer models employ self-attention mechanisms to process input sequences in parallel, capturing long-range dependencies. Encoders generate contextual representations, while decoders generate output sequences. Key components include input/positional encoding, multi-head attention, feed-forward neural networks, and layer normalization. 
 }
\label{fig:transformer}
\end{figure*}

\paragraph{Transformers}
Transformers represent a groundbreaking deep learning architecture that has revolutionized natural language processing. Departing from traditional Recurrent Neural Networks (\texttt{RNN}s) \citep{rumelhart1986learning}, transformers employ an attention mechanism to weigh the significance of input sequence elements. This approach enables parallel processing, substantially enhancing computational efficiency \citep{EfficientTransformers}. By encoding input sequences into dense vector representations and leveraging self-attention to capture intricate inter-word relationships, transformers have achieved state-of-the-art performance in tasks such as machine translation, text summarization, and question answering. Their capacity to handle long-range dependencies and learn contextual representations has expanded their application to domains beyond natural language, including computer vision and audio processing. \cref{fig:transformer} illustrates the basic transformer model.

\paragraph{Long Short-Term Memory (\texttt{LSTM})}

\texttt{LSTM} \citep{hochreiter1997long} is a type of recurrent neural network (\texttt{RNN}) designed to address the vanishing gradient problem. \texttt{LSTM}s excel at handling sequential data by employing a complex cell structure with input, output, and forget gates. These gates regulate the flow of information, enabling the network to selectively remember or forget information over extended periods, capturing long-term dependencies effectively \citep{lstm_survey}.

\paragraph{Diffusion models}
Diffusion models are generative models that gradually add noise to training data and then learn to reverse this process. They excel at generating high-quality, diverse samples by modeling the data distribution implicitly. Key components include a forward diffusion process, a reverse diffusion process learned through neural networks, and a sampling procedure to generate new data \citep{yang2023diffsurvey}.

\paragraph{Generative Adversarial Networks (\texttt{GAN}s)}
\texttt{GAN}s are a class of machine learning frameworks pitting two neural networks against each other in a competitive process. A generator creates new data instances, while a discriminator evaluates their authenticity. Through adversarial training, the generator learns to produce increasingly realistic outputs that deceive the discriminator \citep{GAN}.

\subsubsection{Retrievers}
Retrievers are essential components of information retrieval systems, tasked with efficiently locating relevant information within vast datasets. By transforming both queries and documents into dense vector representations, retrievers can rapidly identify and rank documents based on their semantic similarity to the given query. This process, often leveraging techniques like embedding and similarity search, forms the foundation for many modern search and question-answering applications \citep{zamani2022retrieval}. 

\paragraph{Sparse Retrievers}
Sparse Retrievers are traditional information retrieval methods that rely on lexical matching to find relevant documents. They represent documents and queries as sparse vectors, typically using techniques like \texttt{TF-IDF} \citep{DBLP:conf/sigir/RobertsonW97} or \texttt{BM25} \citep{DBLP:conf/sigir/LaffertyZ01}. These models excel at capturing term frequencies and inverse document frequencies, allowing for efficient search and ranking based on exact keyword matches. While effective for certain tasks, sparse retrievers often struggle to capture semantic meaning and handle complex query formulations \citep{DBLP:journals/ftir/RobertsonZ09}.

\paragraph{Dense Retrievers}
Dense retrievers employ deep learning to project textual data into high-dimensional semantic spaces. By learning distributed representations that capture contextual and semantic information, these models enable efficient similarity search and retrieval. Architectures such as \texttt{BERT} and its derivatives have significantly advanced the field, demonstrating superior performance in capturing complex linguistic patterns compared to traditional sparse retrieval methods based on lexical matching \citep{DBLP:conf/iclr/GuoRLFT0ZDSFTDC21}.

\subsubsection{\rag~foundation}

The foundation of a \rag~pipeline is vector storage. This process involves several key steps, each crucial for creating an effective system for semantic search and advanced \nlp~tasks. The key components are

\paragraph{Document Ingestion}
Document loaders serve as versatile tools for handling various data formats:
\begin{itemize}
    \item Process multiple file types (e.g., PDF, JSON, HTML, Markdown)
    \item Direct access to databases and \texttt{API}s (e.g., \texttt{GitHub}, \texttt{Reddit}, \texttt{Google Drive})
\end{itemize}

\paragraph{Document Segmentation}
Text splitters break down documents into smaller, meaningful chunks:
\begin{itemize}
    \item Employ intelligent splitting methods over simple character-based division
    \item Examples include splitting by header or using recursive techniques
    \item Retain relevant metadata during the splitting process
\end{itemize}

\paragraph{Vector Embedding}
Embedding models transform text into vector representations:
\begin{itemize}
    \item Enable nuanced language understanding
    \item Facilitate effective semantic search capabilities
\end{itemize}

\paragraph{Vector Storage}
Specialized vector databases optimize storage and retrieval:
\begin{itemize}
    \item Examples include \texttt{Chroma}\footnote{\url{https://www.trychroma.com/}}, \texttt{Pinecone}\footnote{\url{https://www.pinecone.io/}}, \texttt{Milvus} \footnote{\url{https://milvus.io/}}, \texttt{FAISS} \footnote{\url{https://faiss.ai/}}, and \texttt{Annoy} \footnote{\url{https://github.com/spotify/annoy}}
    \item Enable efficient retrieval based on vector similarity
\end{itemize}

\subsection{Overview on \llava: Large Language and Vision Assistant}

\llava~is a multimodal \ai~model that integrates natural language processing and computer vision. It processes and reasons about both text and images by combining a vision encoder with a large language model \citep{liu2024visual}. 

\paragraph{Architecture}
LLaVA employs a pretrained vision transformer (\texttt{ViT}) \citep{dosovitskiy2020image} to encode images. This visual information is integrated with a large language model (e.g., \llama) through a projection layer.

\paragraph{Training}
\llava~undergoes a two-phase training process to enhance its ability to comprehend user instructions and generate accurate responses based on both textual and visual input. 
\begin{itemize}
    \item Pre-training for Feature Alignment: This initial phase aligns visual and language features for compatibility. 
    \item Fine-tuning End-to-End: The entire model is refined in this stage. While the vision encoder's weights remain fixed, the projection layer and language model parameters are adjusted. This adaptable process enables \llava~to excel in various applications. 
\end{itemize}

\llava~surpasses its predecessor by incorporating specialized visual question answering datasets aimed at academic research. These datasets are tailored to tasks involving visual question answering , optical character recognition, and region-level analysis. This advancement enhances \llava's capabilities, enabling it to excel in applications such as text recognition and the accurate localization of detailed visual elements. It is worth to highlight that \llava~has shown promising on various dataset such as science question answering \citep{lu2022learn} and outperformed models like \texttt{Blip-2}\citep{li2023blip} and \citep{awadalla2023openflamingo}. In particular, when fine-tuned on question answering, \llava~in synergy with \gpt-4 achieves 92.53\%  accuracy \citep{liu2024visual}.

\subsection{Overview on Stable Diffusion Models (\sdm s) for image generation}

\sdm s represent a significant advancement in the field of generative models for image synthesis, drawing from principles in both probabilistic modeling and deep learning \citep{zhang2023adding}. These models are built upon the concept of diffusion processes, which are mathematical models used to describe the gradual transition of a system from one state to another, often in the context of noise and information diffusion.

\subsubsection{Diffusion Models: Fundamentals}

Diffusion models generate images through an iterative process of noise reduction. Starting from a random, noisy image, the model progressively refines it into a realistic image by applying learned denoising steps. This process is inspired by the gradual diffusion of ink in water, where a clear image emerges over time \citep{yang2023diffusion}.

\begin{itemize}
    \item Forward Process: Initially, Gaussian noise is systematically added to an image in multiple steps, transforming it into a nearly unrecognizable noise pattern. This process is deterministic.
    \item Reverse Process: The model learns to reverse the noise addition by predicting how to remove noise at each step. This is accomplished through a neural network trained to reconstruct the original image from its noisy version.
    \item Training Objective: The model's goal is to accurately predict the noise that needs to be removed at every step. This is achieved by minimizing the difference between the model's predicted denoised image and the ground truth clean image. Techniques like variational inference or score matching are commonly used for training.
\end{itemize}

\subsubsection{Importance and Applications of \sdm s}

\begin{itemize}
    \item High-Quality Image Generation: Stable diffusion models can produce high-resolution, photorealistic images. This capability is particularly important for applications requiring high fidelity and detail, such as digital art, virtual reality, and simulation.

    \item Controlled Image Synthesis: These models can generate images based on specific conditions or prompts. This controlled synthesis capability is valuable for tasks like conditional image generation, where images need to adhere to certain attributes or themes.

    \item Robustness and Stability: Compared to some other generative models, such as \texttt{GAN} \citep{goodfellow2014generative}, diffusion models exhibit greater stability during training. \texttt{GAN}s can suffer from issues like mode collapse or instability \citep{thanh2020catastrophic}, whereas diffusion models benefit from a more straightforward and stable training process \citep{li2024generalization}.

    \item Flexibility: Diffusion models can be adapted for various tasks beyond image generation, including text-to-image synthesis, in painting (i.e., image completion), and style transfer. Their flexibility makes them a versatile tool in the field of machine learning and computer vision.

    \item Theoretical Insights: Diffusion models contribute to a deeper understanding of generative processes. They bridge probabilistic modeling with neural network training, offering insights into how noise and information can be managed and utilized in image synthesis.
    
\end{itemize}

\subsection{Overview on Low-Rank Adaptation (\lora) Weights}
\lora\ is a technique designed to improve the efficiency of fine-tuning large models by focusing on a reduced number of parameters, which are learned in low-rank subspaces \citep{hu2021lora}. In the realm of advanced generative models, \lora\ weights, represent a pivotal enhancement for Stable Diffusion models. \lora\ weights are designed to optimize the efficiency and performance of large-scale neural networks by adapting them with low-rank matrices. This approach allows for the fine-tuning of complex models with reduced computational overhead and memory usage. By leveraging \lora~weights, users can achieve high-quality image generation and more precise control over the model's outputs without the need for extensive computational resources. This scientific refinement is particularly beneficial in educational and practical applications, where resource constraints are often a consideration. The application of \lora\ weights not only enhances the model's scalability and efficiency but also ensures that advanced generative capabilities remain accessible and practical for a wide range of uses.

\subsection{ByteDance/SDXL-Lightning (\bsdxl)}

Stable Diffusion XL (\sdxl) represents a significant advancement in text-to-image generation, building upon its predecessor \sdm. It introduces three major innovations \footnote{
\href{https://huggingface.co/docs/diffusers/en/using-diffusers/sdxl}{https://huggingface.co/docs/diffusers/en/using-diffusers/sdxl}, \href{https://stability.ai/stable-image}{https://stability.ai/stable-image}}:
\begin{enumerate}
    \item Enhanced Model Architecture: The UNet backbone, a convolutional neural network architecture specifically designed for image segmentation tasks, is tripled in size, substantially increasing processing capacity; It incorporates a dual text encoder system:
     (i) retains the original text encoder
     (ii) integrates \texttt{OpenCLIP ViT-bigG/14} \citep{schuhmann2022laionb, Radford2021LearningTV} as a second encoder. This dual-encoder approach dramatically expands the model's parameter count, enabling more nuanced text understanding and image generation.

 \item Advanced Conditioning Techniques:
   (i) implements novel size and crop-conditioning mechanisms;
   (ii) these techniques preserve valuable training data that might otherwise be discarded;
   (iii) offers users greater control over image cropping and composition in the generation process.

\item Two-Stage Generation Pipeline:
   (i) introduces a base model and a refiner model, working in tandem where the base model generates an initial image (also capable of standalone operation) and the refiner model then enhances this image, adding intricate, high-quality details;
   (ii) this two-stage approach allows for more refined and detailed output.
\end{enumerate}
These enhancements collectively contribute to \sdxl~improved image quality, more accurate text-to-image correspondence, and increased user control over the generation process.

\subsection{Overview on quantization}
Quantization is the process of reducing the precision of the numerical representations used in a model. This typically involves: (i) Converting weights and activations from high-precision formats (e.g., 32-\texttt{bit} floating-point) to lower-precision formats (e.g., 8-\texttt{bit} integers) (ii) Mapping a wide range of values to a smaller set of discrete values. Lower-precision data types occupy less memory, allowing larger models to fit into limited hardware resources. In addition, operations on lower-precision data types are generally faster, leading to quicker model execution. Finally, reduced computational complexity often results in lower power consumption, which is particularly beneficial for edge devices and mobile applications.

The Transformers library by \texttt{Hugging Face} \footnote{\url{https://huggingface.co/docs/transformers/en/main_classes/quantization}} supports several quantization methods: (i) Activation-aware Weight Quantization (AWQ): A technique that considers activation statistics during the quantization process, (ii) GPT Quantization (GPTQ): An algorithm specifically designed for quantizing large language models like \gpt; (iii) Bitsandbytes Quantization: 
8-\texttt{bit} quantization: Reduces precision to 8 \texttt{bits}, 4-\texttt{bit} quantization: Further reduces precision to 4 \texttt{bits}, offering even greater memory savings.

While quantization offers numerous advantages, it's important to note the tradeoffs: (i) there may be a slight decrease in model accuracy, though modern techniques often minimize this impact; (ii) the effectiveness of quantization can vary depending on the specific model architecture and task; (iii) some hardware accelerators are specifically optimized for lower-precision operations, potentially offering additional performance gains \citep{proskurina2024quantization, gong2024makes}.

By leveraging these quantization techniques, developers can deploy larger and more complex models in resource constrained environments, opening up new possibilities for \ai~applications across various domains.

\subsection{Overview on \texttt{Whisper-Base.en}: A Transformer-Based Model for Robust English Speech Recognition}
Whisper-Base.en is a member of the \texttt{Whisper} series of models developed by \openai, designed specifically for automatic speech recognition (\asr) in English. Here we present an overview of Whisper-Base.en, detailing its architectural innovations, training methodologies, and performance characteristics. We discuss its significance in the context of modern speech recognition systems, highlighting its robustness, versatility, and practical applications \citep{radford2023robust}.

\asr\ has seen substantial advancements with the advent of deep learning technologies. Among these advancements, the \texttt{Whisper} series by \openai~represents a notable development, leveraging transformer architectures to address various challenges in speech recognition. \texttt{Whisper-Base.en} is a specific variant tailored for English, exhibiting strong performance in diverse and challenging auditory environments.

\texttt{Whisper-Base.en} employs a transformer-based architecture, which has become the de facto standard for sequence-to-sequence tasks due to its ability to handle long-range dependencies and complex temporal relationships. The model consists of an encoder-decoder framework, where:

\begin{itemize}
    \item Encoder: Processes the raw audio signal by extracting features through a series of self-attention mechanisms and feed-forward layers. This encoder is designed to handle varying lengths of input audio sequences and capture nuanced phonetic details.
  
    \item Decoder: Generates the text output from the encoded representations. The decoder uses attention mechanisms to focus on relevant parts of the encoded audio, facilitating accurate transcription.
\end{itemize}
This architecture enables \texttt{Whisper-Base.en} to effectively model the sequential nature of speech, making it capable of producing coherent and contextually accurate text. It is trained using a combination of supervised learning and extensive datasets, incorporating diverse English accents, dialects, and background noise conditions \footnote{\url{https://huggingface.co/openai/whisper-base.en}}.

\section{\bioeng: Multimodal \cbot~for research and educational purposes}

\begin{figure*}[!htb]
    \centering
    \scalebox{2}{\includegraphics[width=3in]{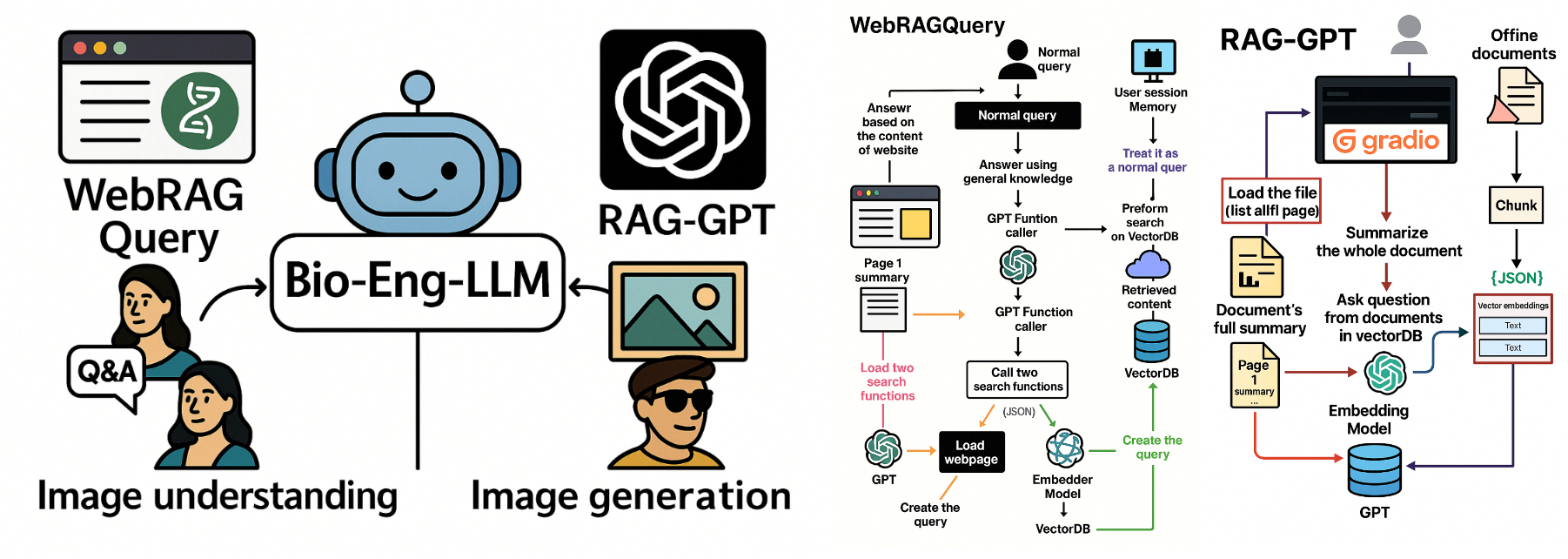}} 
    \caption{Illustration of Multimodal \cbot~ with the capability of image understanding, image generation, document summary, question/answering from user and internet search, and automatic speech recognition.}
    \label{fig:schema}
\end{figure*}

\begin{figure*}[!htb]
\centering
\includegraphics[width=0.99\linewidth]{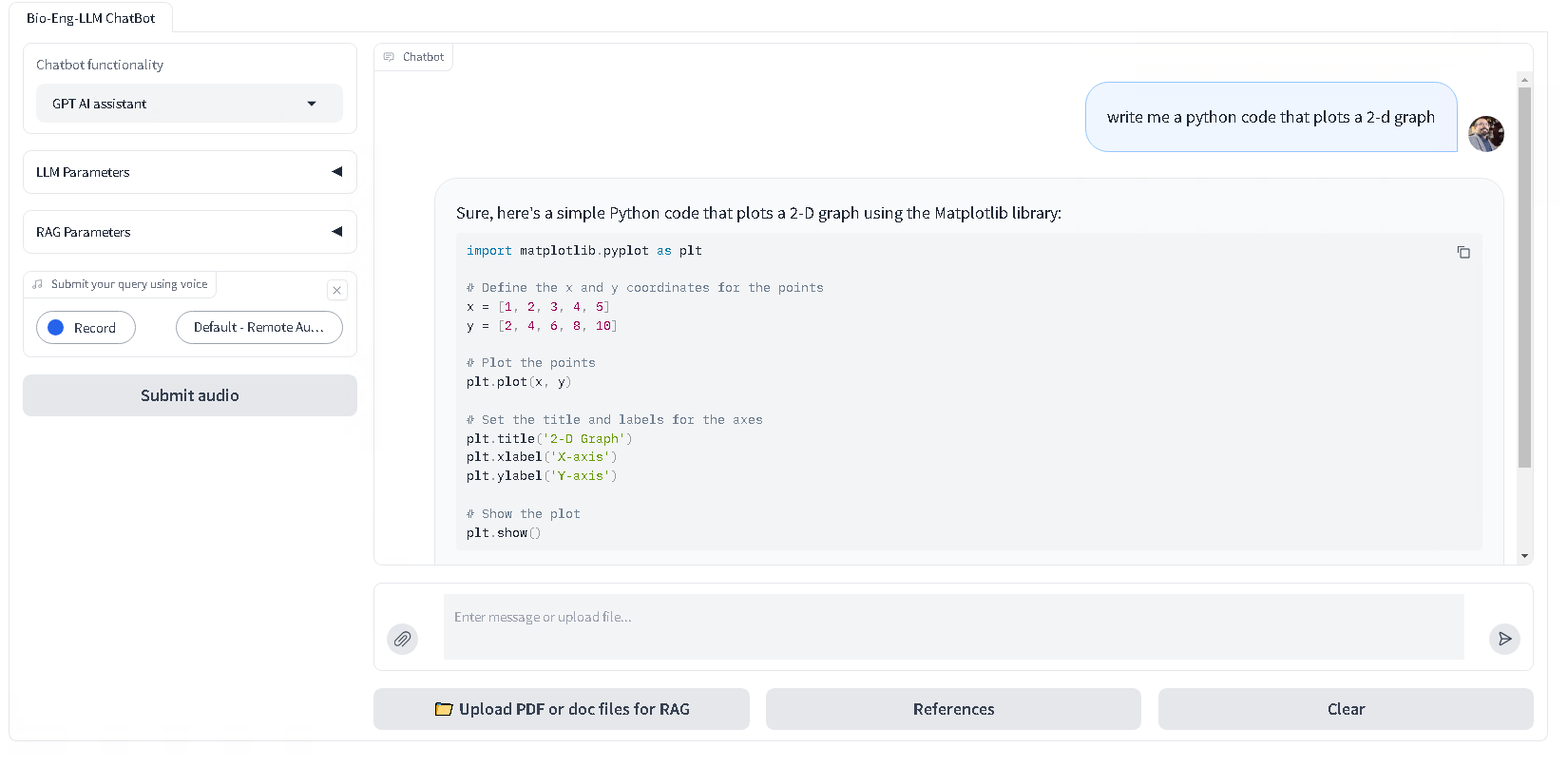}
\caption{\bioeng~\cbot: responding to the user request for writing \texttt{Python} code.}
\label{fig:overview2}
\end{figure*}

\bioeng~implements a graphical user interface (\texttt{GUI}) for a sophisticated \ai-powered \cbot~bot system using the \texttt{Gradio} \texttt{Python} library. The application integrates various advanced natural language processing and computer vision technologies, including \gpt-based language models, \llava~for image understanding, \rag~for enhanced question-answering, and web-based information retrieval. The interface offers multiple functionalities such as general \ai~assistance, image analysis, document summarization, and even image generation using stable diffusion models. The \bioeng~structure suggests a modular design, with separate utility classes for handling chat functionality, file uploads, and \texttt{UI} settings. This multifaceted approach allows users to interact with cutting-edge \ai~models through a user-friendly interface, enabling tasks ranging from simple conversations to complex document analysis and multimodal interactions. In \cref{fig:schema} we illustrate the core idea of \bioeng. \cref{fig:overview2} shows the \texttt{GUI} of the \bioeng~while we asked the \gpt~model to provides a \texttt{Python} code to plot a 2-D graph. In the following subsections we explain different functionalities that we consider in \bioeng.

\subsection{\rag~functionality}

\bioeng~implementation utilizes a sophisticated \rag~system integrated within a multi-functional conversational \ai~framework. The \rag~component employs dense vector representations and semantic search techniques to enhance natural language understanding and generation. It utilizes both pre-indexed corpora and dynamic, user-supplied document collections, stored in optimized vector databases for efficient information retrieval.
At the heart of the \rag~is vector database which utilizes \openai-generated embeddings for efficient document indexing and retrieval. The system is designed to process a collection of textual documents, specifically PDFs, by loading them, segmenting the content into smaller, more manageable text units (or ``chunks"), and embedding these segments into vector space for downstream tasks such as semantic search. The embedded text chunks are then transformed into high-dimensional vectors using \openai~embeddings, which capture the underlying semantic relationships within the text. These vector representations are stored in a ``\texttt{Chroma}" vector database, a high-performance system optimized for handling large-scale embeddings and supporting similarity-based queries. By persisting the VectorDB, the system enables sophisticated information retrieval, allowing for queries to be matched with semantically similar text fragments from large document collections. This approach is highly applicable in fields that require large-scale document analysis, such as computational linguistics, legal document processing, and scientific literature mining, where it significantly improves the speed and accuracy of text retrieval by leveraging the power of vector embeddings.

The system's architecture incorporates parameterizable retrieval mechanisms, including top-k retrieval \citep{ilyas2008survey} and Maximum Marginal Relevance (\texttt{MMR}) \citep{carbonell1998use}, to balance relevance and diversity in retrieved context. \texttt{MMR} aims to select documents that are not only relevant to the query but also dissimilar to each other. This helps to avoid redundancy in the results and provide a more comprehensive and informative response \citep{xia2015learning}.

The \rag~pipeline integrates seamlessly with \llm~for context-aware response generation, leveraging techniques such as prompt engineering and temperature-controlled sampling. The \bioeng~modular design allows for the incorporation of various embedding models and \llm~architectures, facilitating adaptability to different domains and tasks. Furthermore, the implementation features a hybrid approach that combines local knowledge bases with real-time web information retrieval, enabling dynamic knowledge augmentation.

\bioeng~\rag~system exemplifies recent advancements in hybrid \ai~models that synergistically combine information retrieval and natural language generation. By contextualizing \llm~outputs with retrieved information, it addresses challenges such as hallucination \citep{ji2023survey} and knowledge cutoff, while potentially improving the interpretability and factual grounding of generated responses. knowledge cutoff refers to the date after which the model has no awareness of events, facts, or information. The model is trained on data available up to a certain point in time, and anything that happens after this cutoff is not included in the model's knowledge base. The integration of this \rag~component within a larger conversational \ai~framework demonstrates the practical application of state-of-the-art techniques in natural language processing, information retrieval, and machine learning to create more robust and informative dialogue systems.

In the following subsections we describe significant functionalities of the developed \rag~system.
\subsubsection{Web search functionality}
To enhance the \bioeng~information retrieval capabilities, a dedicated search module was integrated. This component leverages the \duck~\footnote{DuckDuckGo is an American software company that offers a number of products intended to help people protect their online privacy \url{https://duckduckgo.com/}.} search API to efficiently extract relevant data from the web. The application provides a structured method for querying diverse content formats, including text, PDFs, images, videos, maps, news, and instant answers. Additionally, the implementation of instant answer retrieval and search suggestion features augments the user experience, facilitating rapid access to crucial information.

\subsubsection{Web summarization functionality}
A web summarization functionality is designed to be integrated within \bioeng, leveraging the \openai~\gpt~engine to process and summarize webpages. Notably, it handles the complexities of webpage content length by segmenting the webpage into smaller sections and iteratively summarizing them using the \llm. The module then generates a final summary by feeding the concatenated summaries from each section back into the \llm. This approach addresses the limitations of \llm s in handling very long inputs while enabling the system to process and summarize entire webpages. Overall, this module implementation demonstrates a strategy for incorporating \llm s into software systems for web content summarization.

\begin{figure*}[!htb]
\centering
\includegraphics[width=0.99\linewidth]{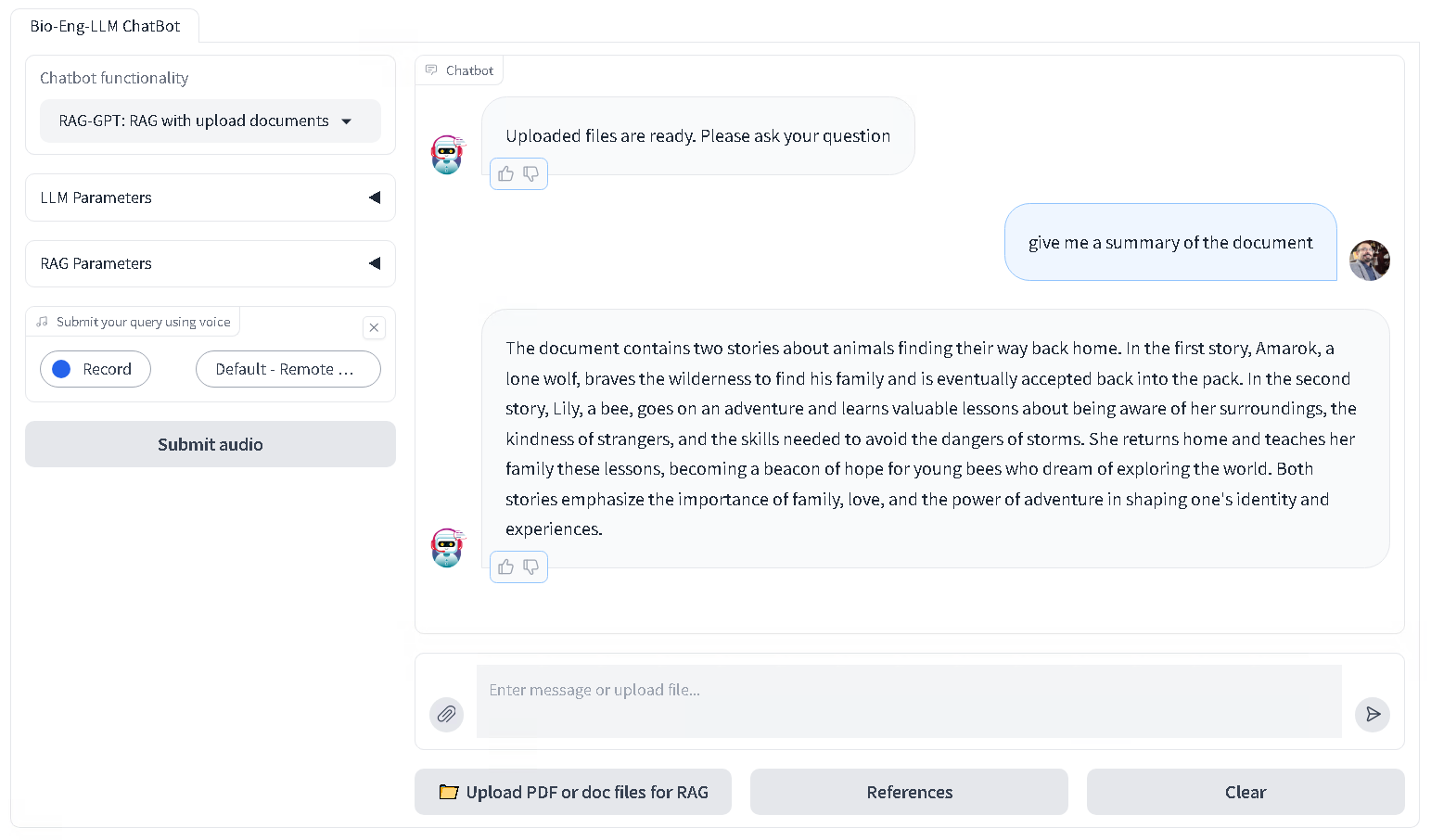}
\caption{\bioeng~\cbot: \rag~functionality on document summarization. Here a pdf document is uploaded and the user asks the summary of it.}
\label{fig:overview_document_summary}
\end{figure*}

\subsection{Image understanding functionality}

\bioeng~implements a sophisticated web service for multimodal \ai~interaction, exploiting the \llava~model. The system utilizes state-of-the-art transformer architectures from the \texttt{Hugging Face}\footnote{\url{https://huggingface.co/llava-hf}} library, integrating both language and vision capabilities. It employs quantization techniques, such as 4-\texttt{bit} quantization with the \texttt{BitsAndBytesConfig} \footnote{\url{https://huggingface.co/docs/transformers/en/main_classes/quantization}}, to optimize memory usage and inference speed on \texttt{GPU} hardware.

The implementation features a \texttt{Flask}-based \texttt{API} endpoint that processes requests containing textual prompts and image \texttt{URL}s. The service dynamically loads and processes images, combining them with text inputs to generate contextually relevant responses. The system supports both quantized and full-precision models, adapting to available computational resources. It implements efficient memory management practices, including \texttt{CUDA} memory clearing, to handle potential resource constraints in production environments.

\begin{figure*}[!h]
\centering
\includegraphics[width=0.99\linewidth]{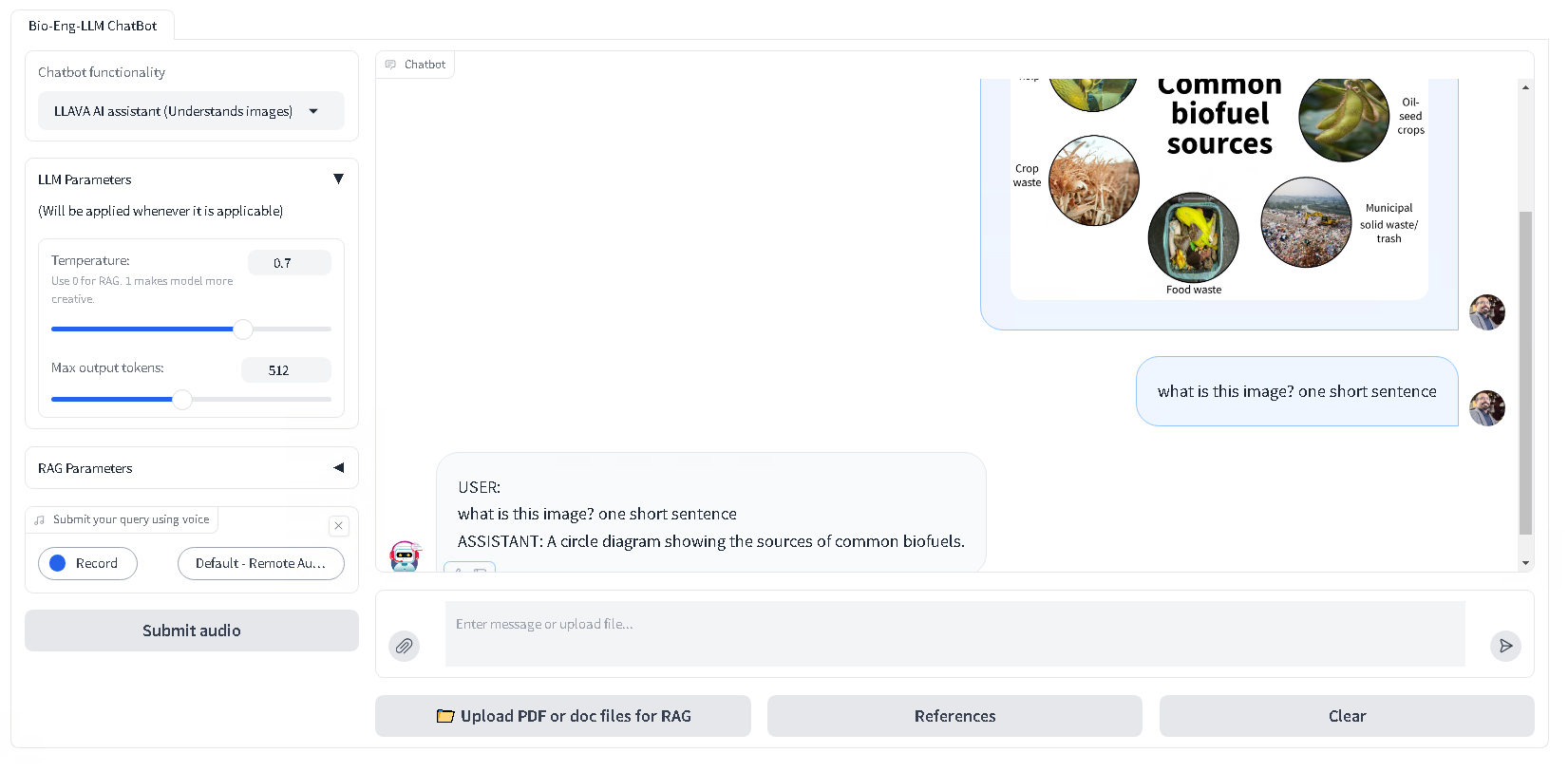}
\caption{\bioeng~\cbot: response of \bioeng~\cbot~to an uploaded image related to bio-fuel sources.}
\label{fig:overview_image_understanding}
\end{figure*}

The \bioeng~architecture allows for flexible deployment, supporting various model sizes and quantization levels to balance performance and resource utilization. By providing a restful API interface, this implementation facilitates the integration of advanced \ai~capabilities into broader applications, enabling complex natural language understanding and generation tasks that incorporate visual context. This approach represents a significant step towards more comprehensive and context-aware \ai~systems capable of processing and reasoning across multiple modalities. In figure \ref{fig:overview_image_understanding} the response of \bioeng-\llm to an uploaded image related to bio-fuel sources is shown.

\subsection{Image generation functionality}
\begin{figure*}[!h]
\centering
\includegraphics[width=0.99\linewidth]{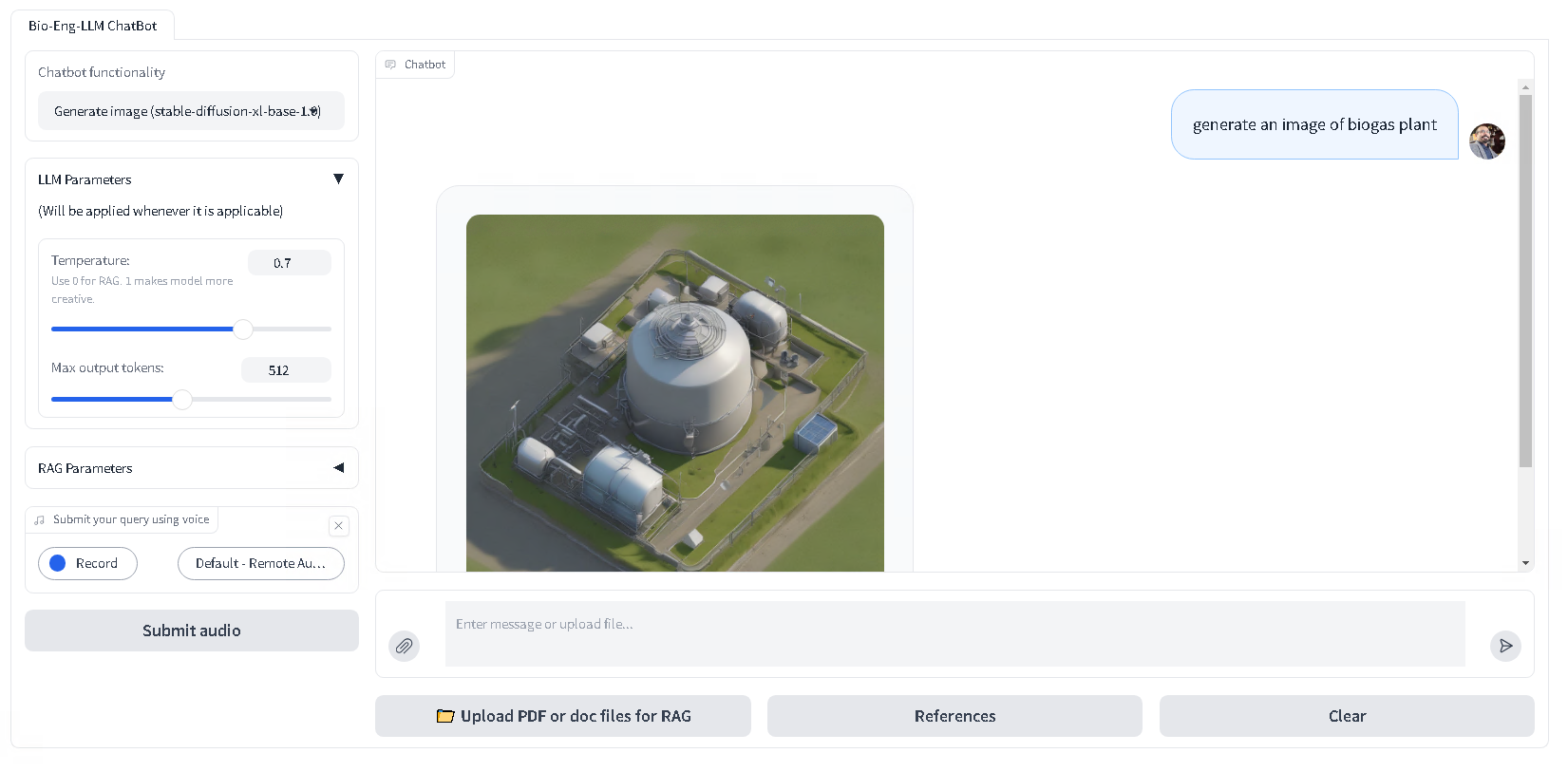}
\caption{\bioeng~\cbot: functionality of \bioeng~on image generation. Here, to generate an image of a biogas plant.}
\label{fig_overview_image_generation}
\end{figure*}

For on-demand image synthesis, \bioeng~exploits the \sdxl~model. The system's architecture employs \texttt{PyTorch} for \texttt{GPU}-accelerated inference, utilizing \texttt{CUDA} when available, and incorporates advanced optimization techniques such as \lora~for efficient fine-tuning. The \sdxl~pipeline is configured with float point $16$ precision and an Euler Discrete Scheduler with trailing \texttt{timesteps}, optimizing for both performance and quality \citep{karras2022elucidating}\footnote{The Euler Discrete Scheduler (EDS) is a numerical approximation method rooted in the theory of ordinary differential equations (ODEs). It is applied in the context of diffusion models, a generative modeling framework that iteratively denoises random noise to produce structured data, such as images. The EDS discretizes the continuous diffusion process into finite time steps, enabling computational tractability\citep{kloeden1992stochastic}}. The \texttt{API} endpoint, orchestrates the image generation process by interpreting \texttt{JSON}-formatted prompts, executing the diffusion process with a notably low inference step count of $4$ and a guidance scale of $0$, and persisting the resultant image. \bioeng~implementation showcases the harmonization of high-performance computing with accessible web services, though it presents opportunities for enhancement in areas such as error handling, input validation, and scalability. \bioeng's design reflects a nuanced understanding of both the technical intricacies of diffusion models and the practical considerations of deploying such models in a production environment. In \cref{fig_overview_image_generation} the response of \bioeng~ to the user's prompt for image generation of biogas plant is shown.

\subsection{Automatic Speech Recognition}

\bioeng~ demonstrates a sophisticated integration of the \texttt{Whisper} \asr~model within a restful \texttt{API} framework. The implementation leverages the \texttt{Hugging Face} Transformers library to instantiate a pipeline for the model, optimizing for English language transcription tasks. 

The service exposes a endpoint that processes post requests containing \texttt{JSON}-formatted audio data, comprising the sampling rate and raw audio samples as a float array \footnote{https://www.json.org/json-en.html}. The audio preprocessing pipeline incorporates normalization techniques, scaling the input to a peak amplitude of 1.0 to ensure consistent model performance across varying input volumes. The \asr~pipeline is elegantly integrated into the Flask route handler, seamlessly converting the preprocessed audio into text. \bioeng~architecture showcases an efficient approach to deploying state-of-the-art speech recognition capabilities as a microservice, although it presents opportunities for enhancement in areas such as error handling, input validation, and potential batch processing capabilities. The system's design reflects a nuanced understanding of both \asr~model deployment and web service architecture, providing a scalable foundation for speech-to-text applications while maintaining a focus on simplicity and performance.

\subsection{Source code availability and hardware specifications}

To develop \bioeng, we utilize a robust and high-performance computing environment. Our setup includes:

\begin{itemize}

    \item Operating System: \texttt{Windows} 2019, 64 \texttt{bit} 

    \item Hardware:
    \begin{itemize}
       
        \item \texttt{GPU} Support: \nvidia~Tesla V100 with 32 \texttt{GB} VRAM
        \item \texttt{RAM}: 768~\texttt{GByte}
        \item \texttt{CPU}: \intel~\texttt{XEON} Platinum 8280 / 4 * 28 Core
        
    \end{itemize}
 This powerful configuration allows us to efficiently process large-scale biological data and run complex machine learning models.

\item Software:
We exclusively use officially released versions of packages in \bioeng~to ensure stability and reproducibility. Key components include:

\begin{itemize}
    \item \llm: Various \llm~models are integrated to \bioeng's functionality. These models are sourced from \texttt{Hugging Face}\footnote{\url{https://huggingface.co/models}}, a leading platform for machine learning models. When selecting models, we carefully consider their terms of use to ensure compliance with licensing requirements.

    \item Package Management:
To maintain a consistent development environment across our team, we use \texttt{Conda}. This approach allows us to: (i)
 Specify exact versions of dependencies, (ii) Easily replicate the environment on different machines, and (iii) Manage potential conflicts between packages

\item Version Control:
We use \texttt{Git} to track changes in our codebase, facilitating collaboration and maintaining a history of our development process.

\end{itemize}

\end{itemize}

\bioeng~made upon \texttt{Python} environment containing a collection of libraries essential for sophisticated natural language processing and machine learning tasks. Fundamental components encompass \texttt{PyTorch}, a deep learning framework,
 along with its supplementary tools \texttt{torchvision} and \texttt{torchaudio} \footnote{\url{https://pytorch.org/}}. \texttt{Hugging Face's Transformers}\footnote{\url{https://huggingface.co/docs/transformers/installation}} library is integrated for cutting-edge natural language comprehension and generation models. \texttt{Accelerate}\footnote{\url{https://huggingface.co/docs/accelerate/en/basic_tutorials/install}}, a high-performance deep learning library, is incorporated for optimized training and inference processes. Complementary packages such as \texttt{Langchain}\footnote{\url{https://python.langchain.com/v0.1/docs/get_started/installation/}}, \texttt{Langchain-Community}, and \texttt{OpenAI}\footnote{\url{https://platform.openai.com/docs/quickstart}} offer foundational utilities for developing language models and interacting with language model \texttt{APIs}. For effective model deployment, \texttt{Uvicorn}\footnote{\url{https://www.uvicorn.org/}} and \texttt{Gradio}\footnote{\url{https://www.gradio.app/guides/quickstart}} are included. To support these functionalities, auxiliary libraries like \texttt{bitsandbytes}\footnote{\url{https://huggingface.co/docs/bitsandbytes/main/en/installation}}, and \texttt{Tiktoken}\footnote{\url{https://anaconda.org/conda-forge/tiktoken}} are installed. Moreover, \texttt{MKL}\footnote{\url{https://pypi.org/project/mkl/}} is incorporated for potential performance enhancements, and \texttt{Spyder}\footnote{\url{https://docs.spyder-ide.org/3/installation.html}} is added as an integrated development environment tailored for \texttt{Python}.

Note that \bioeng~is publicly available in a \texttt{Github}/\texttt{Gitlab} repositories \footnote{\url{https://github.com/Ali-Forootani/multi_llm/tree/main}, \url{https://git.ufz.de/forootan/ufz_llm}}.

\subsection{Graphic User Interface (\texttt{GUI})} 

\bioeng~make use of \texttt{Gradio} which is a \texttt{Python} library\footnote{\url{https://www.gradio.app/}} that facilitates the rapid development and deployment of user interfaces for machine learning models.
 By abstracting away the complexities of front-end development, it empowers researchers and engineers to create interactive demonstrations and applications that showcase the capabilities of their models. This accelerates the iterative process of model development and evaluation, enabling more efficient exploration of hyperparameters and architectural choices. Furthermore, \texttt{Gradio}'s ability to generate shareable links to deployed applications fosters collaboration and knowledge sharing within the machine learning community.

\section{Enhancing Education through the \bioeng~\ai~\cbot~Platform}

The integration of the \bioeng~\ai~\cbot platform into educational settings represents a significant advancement in pedagogical and research methodologies. The system's core components—AI Assistant, Retrieval Augmented Generation (\rag), and Image Processing—collectively enhance the educational experience through several key mechanisms.

\paragraph{\ai~Assistant} The \ai~Assistant functions as a conversational agent, delivering tailored explanations and support that adapt to individual learning needs. This capability provides on-demand, personalized academic assistance, simulating an interactive tutor and facilitating deeper comprehension of complex subjects.

\paragraph{Retrieval Augmented Generation (\rag)} By leveraging \rag~techniques, the platform integrates information from diverse sources, including preprocessed documents, real-time web data, and user-uploaded files. This integration enhances the quality of responses and supports robust research activities, enabling students to access and synthesize the most current and relevant information efficiently.

\paragraph{Image Processing (Stable Diffusion Model \sdm~and \llava)} The image processing functionalities, powered by Stable Diffusion and \llava, allow for the generation and analysis of visual content. This capability aids in the visualization of complex concepts, historical reconstructions, and artistic projects, thereby enriching the learning experience and fostering creativity.

\paragraph{Internet Search and Summarization (\duck)} Utilizing \duck~for secure web searches and content summarization streamlines the research process. Students benefit from concise, relevant summaries of extensive materials, facilitating efficient information retrieval and synthesis, which is critical for academic research and project development.

\paragraph{Cross-Disciplinary and Skill Development} The platform's versatility supports cross-disciplinary projects, promoting collaborative learning across various fields. It also offers students practical experience with advanced \ai~tools, enhancing their technical skills and preparing them for future careers in AI and related domains.



\section{Implications for Research Purposes}
The use of \llm s in various research fields, such as social science, has been exponentially grown since its introduction. However, the limits of these systems are yet to be discovered.

One significant area of application can be in the knowledge summarization of scientific methods. Through decades scientific methods became more and more complicated, which makes these methods similar to black-box for laypersons. Therefore, using the capabilities of \llm s, researchers may be able to translate and simplify the results of these rigorous quantitative methods to an extent that is understandable. 

\begin{figure}[!htb]
    \centering
    \includegraphics[width=0.85\linewidth]{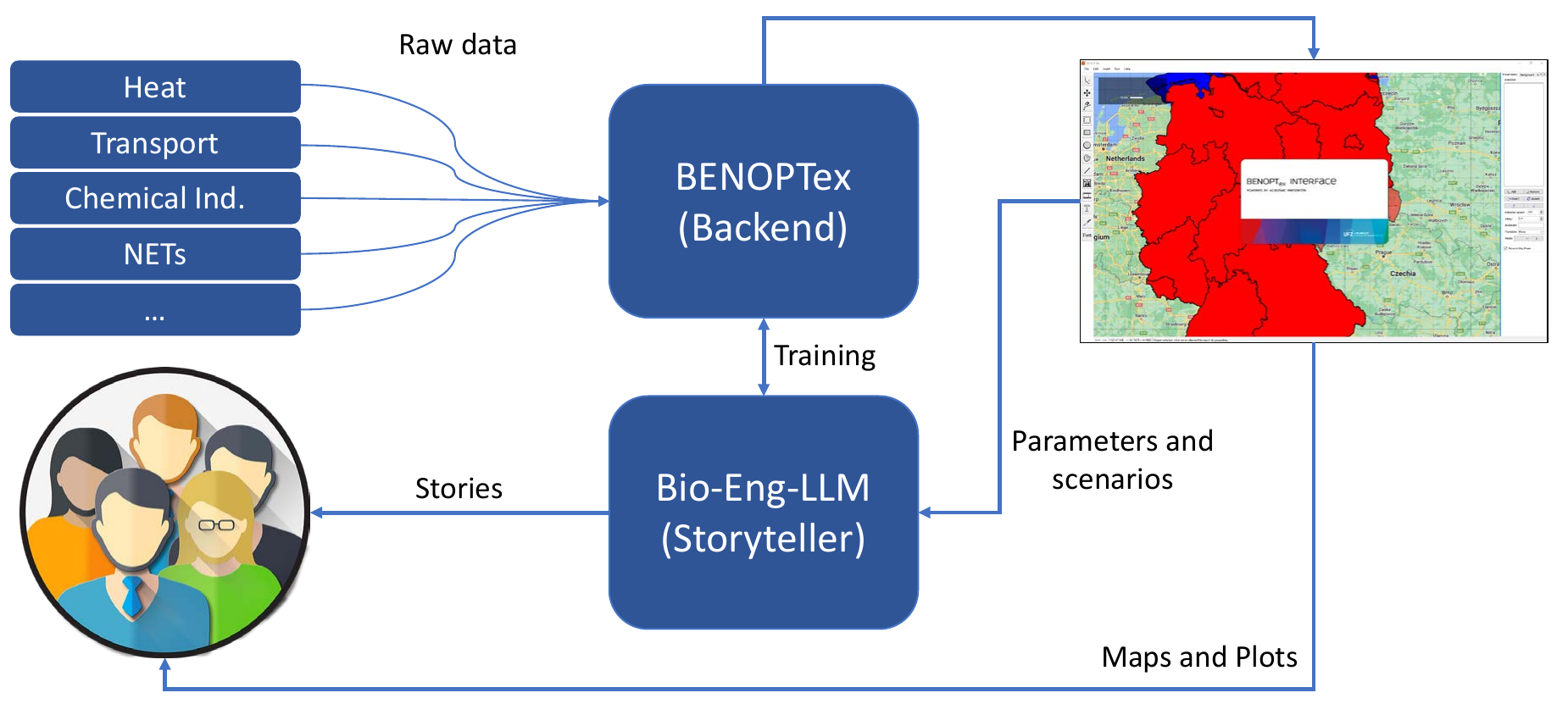}
    \caption{Information flow between  \bioeng~along and \benoptex to create dynamic stories based on users' profile.}
    \label{fig:BENOPTex}
\end{figure}

This approach can assist researchers in closing the loop with stakeholders and converting numerical analyses into stories. We are planning to integrate \bioeng~in the extended bioenergy optimization model (\benoptex) \citep{millinger2022model,esmaeili2023integrating} to translate the outcomes of the optimization models into customized stories such that users can find the implications of these results in their future lives. The users can establish a vivid and coherent dialogue with the \cbot that has been trained based on prior studies, numerical results of the model, and the provided narratives \cite{scenario}. Figure \ref{fig:BENOPTex} shows the schematic of the designed information flow between the \benoptex~ model and \bioeng in order to generate dynamic stories. As depicted in Figure \ref{fig:BENOPTex}, the visualizations on the map are executed in the new BENOPTex interface, which is based on the Academic Presenter platform \citep{avsar2016academic}, while the dialogue takes place in \bioeng.

\section{Conclusion}

In this article we introduced \bioeng~which represents a significant advancement in \ai-assisted educational and research platforms. By integrating cutting-edge technologies such as Retrieval Augmented Generation (\rag), image processing capabilities, and internet search functionalities, this system offers a comprehensive suite of tools for enhancing learning experiences across various disciplines. The platform's ability to generate, understand, and respond to both textual and visual content, coupled with its contextual memory and real-time information retrieval, positions it as a powerful asset in academic settings. \bioeng~not only facilitates creative exploration and technical skill development but also promotes interdisciplinary collaboration, preparing students for the evolving demands of \ai-related fields. As \ai~continues to transform educational paradigms, systems like \bioeng~exemplify the potential for technology to enrich and diversify learning methodologies, ultimately fostering a more dynamic and interactive educational landscape.

The proposed \llm~model in this manuscript is versatile and can be applied to various fields. There are plans to extend and apply the developed tool to energy systems modeling and integrate \bioeng~with \benoptex, benefiting from storytelling techniques. This integration aims to facilitate dynamic interactions with lay users, offering customized insights tailored to their profiles.


\appendix

\section*{CRediT authorship contribution statement}
\textbf{Ali Forootani} Writing – original draft, Visualization, Validation, Software, Resources, Methodology, Formal analysis, Data curation, Conceptualization. \textbf{Danial Esmaeili Aliabadi} Writing – original draft, Visualization, Funding acquisition, Supervision, Conceptualization. \textbf{Daniela Thr\"an} Writing – original draft, Supervision. 

\section*{Funding}
This study is funded by the \href{https://man0euvre.eu/}{\emph{Man0EUvRE}} (100695543), which is co-financed by means of taxation based on the budget adopted by the representatives of the Landtag of Saxony. ‘Man0EUvRE – Energy System Modelling for Transition to a net-Zero 2050 for EU via REPowerEU’ is funded by CETPartnership, the European Partnership under Joint Call 2022 for research proposals, co-funded by the European Commission (GA N°101069750).

\section*{Data availability}
The source code for \bioeng~is publicly available at \url{https://github.com/Ali-Forootani/multi_llm/tree/main} or \url{https://git.ufz.de/forootan/ufz_llm}. The repository contains all the codes needed to run the \bioeng~\cbot.

\bibliographystyle{cas-model2-names}


\bibliography{reference}

\begin{thebibliography}{105}
\expandafter\ifx\csname natexlab\endcsname\relax\def\natexlab#1{#1}\fi
\providecommand{\url}[1]{\texttt{#1}}
\providecommand{\href}[2]{#2}
\providecommand{\path}[1]{#1}
\providecommand{\DOIprefix}{doi:}
\providecommand{\ArXivprefix}{arXiv:}
\providecommand{\URLprefix}{URL: }
\providecommand{\Pubmedprefix}{pmid:}
\providecommand{\doi}[1]{\href{http://dx.doi.org/#1}{\path{#1}}}
\providecommand{\Pubmed}[1]{\href{pmid:#1}{\path{#1}}}
\providecommand{\bibinfo}[2]{#2}
\ifx\xfnm\relax \def\xfnm[#1]{\unskip,\space#1}\fi
\bibitem[{Achiam et~al.(2023)Achiam, Adler, Agarwal, Ahmad, Akkaya, Aleman,
  Almeida, Altenschmidt, Altman, Anadkat et~al.}]{achiam2023gpt}
\bibinfo{author}{Achiam, J.}, \bibinfo{author}{Adler, S.},
  \bibinfo{author}{Agarwal, S.}, \bibinfo{author}{Ahmad, L.},
  \bibinfo{author}{Akkaya, I.}, \bibinfo{author}{Aleman, F.L.},
  \bibinfo{author}{Almeida, D.}, \bibinfo{author}{Altenschmidt, J.},
  \bibinfo{author}{Altman, S.}, \bibinfo{author}{Anadkat, S.}, et~al.,
  \bibinfo{year}{2023}.
\newblock \bibinfo{title}{Gpt-4 technical report}.
\newblock \bibinfo{journal}{arXiv preprint arXiv:2303.08774} .
\bibitem[{Adetayo et~al.(2024)Adetayo, Enamudu, Lawal and
  Odunewu}]{adetayo2024text}
\bibinfo{author}{Adetayo, A.J.}, \bibinfo{author}{Enamudu, A.I.},
  \bibinfo{author}{Lawal, F.M.}, \bibinfo{author}{Odunewu, A.O.},
  \bibinfo{year}{2024}.
\newblock \bibinfo{title}{From text to video with ai: the rise and potential of
  sora in education and libraries}.
\newblock \bibinfo{journal}{Library Hi Tech News} .
\bibitem[{Ahmad et~al.(2021)Ahmad, Chakraborty, Ray and
  Chang}]{ahmad2021unified}
\bibinfo{author}{Ahmad, W.U.}, \bibinfo{author}{Chakraborty, S.},
  \bibinfo{author}{Ray, B.}, \bibinfo{author}{Chang, K.W.},
  \bibinfo{year}{2021}.
\newblock \bibinfo{title}{Unified pre-training for program understanding and
  generation}.
\newblock \bibinfo{journal}{arXiv preprint arXiv:2103.06333} .
\bibitem[{AI(2022)}]{stable_diff}
\bibinfo{author}{AI, S.}, \bibinfo{year}{2022}.
\newblock \bibinfo{title}{stable-diffusion-3-medium}.
\newblock \URLprefix \url{https://huggingface.co/stabilityai}.
\bibitem[{Avsar et~al.(2016)Avsar, Aliabadi, Aliabadi and
  Yousefnezhad}]{avsar2016academic}
\bibinfo{author}{Avsar, B.}, \bibinfo{author}{Aliabadi, D.E.},
  \bibinfo{author}{Aliabadi, E.E.}, \bibinfo{author}{Yousefnezhad, R.},
  \bibinfo{year}{2016}.
\newblock \bibinfo{title}{Academic presenter: A new storytelling presentation
  software for academic purposes}.
\newblock \bibinfo{journal}{arXiv preprint arXiv:1607.06979} .
\bibitem[{Awadalla et~al.(2023)Awadalla, Gao, Gardner, Hessel, Hanafy, Zhu,
  Marathe, Bitton, Gadre, Sagawa et~al.}]{awadalla2023openflamingo}
\bibinfo{author}{Awadalla, A.}, \bibinfo{author}{Gao, I.},
  \bibinfo{author}{Gardner, J.}, \bibinfo{author}{Hessel, J.},
  \bibinfo{author}{Hanafy, Y.}, \bibinfo{author}{Zhu, W.},
  \bibinfo{author}{Marathe, K.}, \bibinfo{author}{Bitton, Y.},
  \bibinfo{author}{Gadre, S.}, \bibinfo{author}{Sagawa, S.}, et~al.,
  \bibinfo{year}{2023}.
\newblock \bibinfo{title}{Openflamingo: An open-source framework for training
  large autoregressive vision-language models}.
\newblock \bibinfo{journal}{arXiv preprint arXiv:2308.01390} .
\bibitem[{Baghdasaryan et~al.(2024)Baghdasaryan, Bunarjyan, Poghosyan,
  Harutyunyan and El-Zein}]{baghdasaryan2024knowledge}
\bibinfo{author}{Baghdasaryan, A.}, \bibinfo{author}{Bunarjyan, T.},
  \bibinfo{author}{Poghosyan, A.}, \bibinfo{author}{Harutyunyan, A.},
  \bibinfo{author}{El-Zein, J.}, \bibinfo{year}{2024}.
\newblock \bibinfo{title}{Knowledge retrieval and diagnostics in cloud services
  with large language models}.
\newblock \bibinfo{journal}{Expert Systems with Applications} ,
  \bibinfo{pages}{124736}.
\bibitem[{Bai et~al.(2024)Bai, Liang, Wan, Yang, Li, Wang, Cui, He, Yuan and
  Zhang}]{bai2024survey}
\bibinfo{author}{Bai, T.}, \bibinfo{author}{Liang, H.}, \bibinfo{author}{Wan,
  B.}, \bibinfo{author}{Yang, L.}, \bibinfo{author}{Li, B.},
  \bibinfo{author}{Wang, Y.}, \bibinfo{author}{Cui, B.}, \bibinfo{author}{He,
  C.}, \bibinfo{author}{Yuan, B.}, \bibinfo{author}{Zhang, W.},
  \bibinfo{year}{2024}.
\newblock \bibinfo{title}{A survey of multimodal large language model from a
  data-centric perspective}.
\newblock \bibinfo{journal}{arXiv preprint arXiv:2405.16640} .
\bibitem[{Bail(2024)}]{bail2024can}
\bibinfo{author}{Bail, C.A.}, \bibinfo{year}{2024}.
\newblock \bibinfo{title}{Can {Generative AI} improve social science?}
\newblock \bibinfo{journal}{Proceedings of the National Academy of Sciences}
  \bibinfo{volume}{121}, \bibinfo{pages}{e2314021121}.
\bibitem[{Borgeaud et~al.(2022)Borgeaud, Mensch, Hoffmann, Cai, Rutherford,
  Millican, Van Den~Driessche, Lespiau, Damoc, Clark
  et~al.}]{borgeaud2022improving}
\bibinfo{author}{Borgeaud, S.}, \bibinfo{author}{Mensch, A.},
  \bibinfo{author}{Hoffmann, J.}, \bibinfo{author}{Cai, T.},
  \bibinfo{author}{Rutherford, E.}, \bibinfo{author}{Millican, K.},
  \bibinfo{author}{Van Den~Driessche, G.B.}, \bibinfo{author}{Lespiau, J.B.},
  \bibinfo{author}{Damoc, B.}, \bibinfo{author}{Clark, A.}, et~al.,
  \bibinfo{year}{2022}.
\newblock \bibinfo{title}{Improving language models by retrieving from
  trillions of tokens}, in: \bibinfo{booktitle}{International conference on
  machine learning}, \bibinfo{organization}{PMLR}. pp.
  \bibinfo{pages}{2206--2240}.
\bibitem[{Brown et~al.(2020)Brown, Mann, Ryder, Subbiah, Kaplan, Dhariwal,
  Neelakantan, Shyam, Sastry, Askell et~al.}]{brown2020language}
\bibinfo{author}{Brown, T.}, \bibinfo{author}{Mann, B.},
  \bibinfo{author}{Ryder, N.}, \bibinfo{author}{Subbiah, M.},
  \bibinfo{author}{Kaplan, J.D.}, \bibinfo{author}{Dhariwal, P.},
  \bibinfo{author}{Neelakantan, A.}, \bibinfo{author}{Shyam, P.},
  \bibinfo{author}{Sastry, G.}, \bibinfo{author}{Askell, A.}, et~al.,
  \bibinfo{year}{2020}.
\newblock \bibinfo{title}{Language models are few-shot learners}.
\newblock \bibinfo{journal}{Advances in neural information processing systems}
  \bibinfo{volume}{33}, \bibinfo{pages}{1877--1901}.
\bibitem[{Cao et~al.(2023)Cao, Li, Liu, Yan, Dai, Yu and
  Sun}]{cao2023comprehensive}
\bibinfo{author}{Cao, Y.}, \bibinfo{author}{Li, S.}, \bibinfo{author}{Liu, Y.},
  \bibinfo{author}{Yan, Z.}, \bibinfo{author}{Dai, Y.}, \bibinfo{author}{Yu,
  P.S.}, \bibinfo{author}{Sun, L.}, \bibinfo{year}{2023}.
\newblock \bibinfo{title}{A comprehensive survey of ai-generated content
  (aigc): A history of generative ai from gan to chatgpt}.
\newblock \bibinfo{journal}{arXiv preprint arXiv:2303.04226} .
\bibitem[{Carbonell and Goldstein(1998)}]{carbonell1998use}
\bibinfo{author}{Carbonell, J.}, \bibinfo{author}{Goldstein, J.},
  \bibinfo{year}{1998}.
\newblock \bibinfo{title}{The use of {MMR}, diversity-based reranking for
  reordering documents and producing summaries}, in:
  \bibinfo{booktitle}{Proceedings of the 21st annual international ACM SIGIR
  conference on Research and development in information retrieval}, pp.
  \bibinfo{pages}{335--336}.
\bibitem[{Chappin(2023)}]{Chappin}
\bibinfo{author}{Chappin, E.}, \bibinfo{year}{2023}.
\newblock \bibinfo{title}{Teaching highly intelligent primary school kids
  energy system complexity}.
\newblock \bibinfo{journal}{Review of Artificial Societies and Social
  Simulation} \URLprefix \url{https://rofasss.org/2023/04/19/teachcomplex}.
  \bibinfo{note}{(accessed on 20 August 2024)}.
\bibitem[{Chen et~al.(2022)Chen, Guo, Yi et~al.}]{DBLP:conf/cvpr/ChenGY0E22}
\bibinfo{author}{Chen, J.}, \bibinfo{author}{Guo, H.}, \bibinfo{author}{Yi,
  K.}, et~al., \bibinfo{year}{2022}.
\newblock \bibinfo{title}{Visualgpt: Data-efficient adaptation of pretrained
  language models for image captioning}, in: \bibinfo{booktitle}{{CVPR}}.
\bibitem[{Chen et~al.(2023a)Chen, Pan, Li, Yao, Chao and
  Mei}]{chen2023retrieval}
\bibinfo{author}{Chen, J.}, \bibinfo{author}{Pan, Y.}, \bibinfo{author}{Li,
  Y.}, \bibinfo{author}{Yao, T.}, \bibinfo{author}{Chao, H.},
  \bibinfo{author}{Mei, T.}, \bibinfo{year}{2023}a.
\newblock \bibinfo{title}{Retrieval augmented convolutional encoder-decoder
  networks for video captioning}.
\newblock \bibinfo{journal}{ACM Transactions on Multimedia Computing,
  Communications and Applications} \bibinfo{volume}{19},
  \bibinfo{pages}{1--24}.
\bibitem[{Chen et~al.(2023b)Chen, Zhu, Shen, Li, Liu, Zhang, Krishnamoorthi,
  Chandra, Xiong and Elhoseiny}]{chen2023minigptv2}
\bibinfo{author}{Chen, J.}, \bibinfo{author}{Zhu, D.}, \bibinfo{author}{Shen,
  X.}, \bibinfo{author}{Li, X.}, \bibinfo{author}{Liu, Z.},
  \bibinfo{author}{Zhang, P.}, \bibinfo{author}{Krishnamoorthi, R.},
  \bibinfo{author}{Chandra, V.}, \bibinfo{author}{Xiong, Y.},
  \bibinfo{author}{Elhoseiny, M.}, \bibinfo{year}{2023}b.
\newblock \bibinfo{title}{Minigpt-v2: large language model as a unified
  interface for vision-language multi-task learning}.
\newblock \bibinfo{journal}{arXiv preprint arXiv:2310.09478} .
\bibitem[{Chen et~al.(2021a)Chen, Tworek, Jun, Yuan, Pinto, Kaplan, Edwards,
  Burda, Joseph, Brockman et~al.}]{chen2021evaluating}
\bibinfo{author}{Chen, M.}, \bibinfo{author}{Tworek, J.}, \bibinfo{author}{Jun,
  H.}, \bibinfo{author}{Yuan, Q.}, \bibinfo{author}{Pinto, H.P.D.O.},
  \bibinfo{author}{Kaplan, J.}, \bibinfo{author}{Edwards, H.},
  \bibinfo{author}{Burda, Y.}, \bibinfo{author}{Joseph, N.},
  \bibinfo{author}{Brockman, G.}, et~al., \bibinfo{year}{2021}a.
\newblock \bibinfo{title}{Evaluating large language models trained on code}.
\newblock \bibinfo{journal}{arXiv preprint arXiv:2107.03374} .
\bibitem[{Chen et~al.(2021b)Chen, Tworek
  et~al.}]{DBLP:journals/corr/abs-2107-03374}
\bibinfo{author}{Chen, M.}, \bibinfo{author}{Tworek, J.}, et~al.,
  \bibinfo{year}{2021}b.
\newblock \bibinfo{title}{Evaluating large language models trained on code}.
\newblock \bibinfo{journal}{arXiv:2107.03374} .
\bibitem[{Chou et~al.(2015)Chou, Chang and Lu}]{chou2015prezi}
\bibinfo{author}{Chou, P.N.}, \bibinfo{author}{Chang, C.C.},
  \bibinfo{author}{Lu, P.F.}, \bibinfo{year}{2015}.
\newblock \bibinfo{title}{Prezi versus powerpoint: The effects of varied
  digital presentation tools on students’ learning performance}.
\newblock \bibinfo{journal}{Computers \& Education} \bibinfo{volume}{91},
  \bibinfo{pages}{73--82}.
\bibitem[{Darvishi et~al.(2024)Darvishi, Khosravi, Sadiq, Ga{\v{s}}evi{\'c} and
  Siemens}]{darvishi2024impact}
\bibinfo{author}{Darvishi, A.}, \bibinfo{author}{Khosravi, H.},
  \bibinfo{author}{Sadiq, S.}, \bibinfo{author}{Ga{\v{s}}evi{\'c}, D.},
  \bibinfo{author}{Siemens, G.}, \bibinfo{year}{2024}.
\newblock \bibinfo{title}{Impact of ai assistance on student agency}.
\newblock \bibinfo{journal}{Computers \& Education} \bibinfo{volume}{210},
  \bibinfo{pages}{104967}.
\bibitem[{Das et~al.(2021)Das, Zaheer, Thai
  et~al.}]{DBLP:conf/emnlp/DasZTGPLTPM21}
\bibinfo{author}{Das, R.}, \bibinfo{author}{Zaheer, M.}, \bibinfo{author}{Thai,
  D.}, et~al., \bibinfo{year}{2021}.
\newblock \bibinfo{title}{Case-based reasoning for natural language queries
  over knowledge bases}, in: \bibinfo{booktitle}{EMNLP}.
\bibitem[{DOSOVITSKIY(2020)}]{dosovitskiy2020image}
\bibinfo{author}{DOSOVITSKIY, A.}, \bibinfo{year}{2020}.
\newblock \bibinfo{title}{An image is worth 16x16 words: Transformers for image
  recognition at scale}.
\newblock \bibinfo{journal}{arXiv preprint arXiv:2010.11929} .
\bibitem[{Esmaeili~Aliabadi et~al.(2023)Esmaeili~Aliabadi, Manske, Seeger,
  Lehneis and Thr{\"a}n}]{esmaeili2023integrating}
\bibinfo{author}{Esmaeili~Aliabadi, D.}, \bibinfo{author}{Manske, D.},
  \bibinfo{author}{Seeger, L.}, \bibinfo{author}{Lehneis, R.},
  \bibinfo{author}{Thr{\"a}n, D.}, \bibinfo{year}{2023}.
\newblock \bibinfo{title}{Integrating knowledge acquisition, visualization, and
  dissemination in energy system models: {BENOPTex} study}.
\newblock \bibinfo{journal}{Energies} \bibinfo{volume}{16},
  \bibinfo{pages}{5113}.
\bibitem[{Gong et~al.(2024)Gong, Liu, Wang, Cai, Zhao and Yan}]{gong2024makes}
\bibinfo{author}{Gong, Z.}, \bibinfo{author}{Liu, J.}, \bibinfo{author}{Wang,
  J.}, \bibinfo{author}{Cai, X.}, \bibinfo{author}{Zhao, D.},
  \bibinfo{author}{Yan, R.}, \bibinfo{year}{2024}.
\newblock \bibinfo{title}{What makes quantization for large language model
  hard? an empirical study from the lens of perturbation}, in:
  \bibinfo{booktitle}{Proceedings of the AAAI Conference on Artificial
  Intelligence}, pp. \bibinfo{pages}{18082--18089}.
\bibitem[{Goodfellow et~al.(2014)Goodfellow, Pouget-Abadie, Mirza, Xu,
  Warde-Farley, Ozair, Courville and Bengio}]{goodfellow2014generative}
\bibinfo{author}{Goodfellow, I.}, \bibinfo{author}{Pouget-Abadie, J.},
  \bibinfo{author}{Mirza, M.}, \bibinfo{author}{Xu, B.},
  \bibinfo{author}{Warde-Farley, D.}, \bibinfo{author}{Ozair, S.},
  \bibinfo{author}{Courville, A.}, \bibinfo{author}{Bengio, Y.},
  \bibinfo{year}{2014}.
\newblock \bibinfo{title}{Generative adversarial nets}.
\newblock \bibinfo{journal}{Advances in neural information processing systems}
  \bibinfo{volume}{27}.
\bibitem[{Goodfellow et~al.(2020)Goodfellow, Pouget-Abadie, Mirza et~al.}]{GAN}
\bibinfo{author}{Goodfellow, I.}, \bibinfo{author}{Pouget-Abadie, J.},
  \bibinfo{author}{Mirza, M.}, et~al., \bibinfo{year}{2020}.
\newblock \bibinfo{title}{Generative adversarial networks}.
\newblock \bibinfo{journal}{CACM} \bibinfo{volume}{63},
  \bibinfo{pages}{139--144}.
\bibitem[{Grossmann et~al.(2023)Grossmann, Feinberg, Parker, Christakis,
  Tetlock and Cunningham}]{grossmann2023ai}
\bibinfo{author}{Grossmann, I.}, \bibinfo{author}{Feinberg, M.},
  \bibinfo{author}{Parker, D.C.}, \bibinfo{author}{Christakis, N.A.},
  \bibinfo{author}{Tetlock, P.E.}, \bibinfo{author}{Cunningham, W.A.},
  \bibinfo{year}{2023}.
\newblock \bibinfo{title}{{AI} and the transformation of social science
  research}.
\newblock \bibinfo{journal}{Science} \bibinfo{volume}{380},
  \bibinfo{pages}{1108--1109}.
\bibitem[{Guo et~al.(2021)Guo, Ren et~al.}]{DBLP:conf/iclr/GuoRLFT0ZDSFTDC21}
\bibinfo{author}{Guo, D.}, \bibinfo{author}{Ren, S.}, et~al.,
  \bibinfo{year}{2021}.
\newblock \bibinfo{title}{Graphcodebert: Pre-training code representations with
  data flow}, in: \bibinfo{booktitle}{ICLR}.
\bibitem[{Guu et~al.(2020)Guu, Lee, Tung, Pasupat and Chang}]{guu2020retrieval}
\bibinfo{author}{Guu, K.}, \bibinfo{author}{Lee, K.}, \bibinfo{author}{Tung,
  Z.}, \bibinfo{author}{Pasupat, P.}, \bibinfo{author}{Chang, M.},
  \bibinfo{year}{2020}.
\newblock \bibinfo{title}{Retrieval augmented language model pre-training}, in:
  \bibinfo{booktitle}{International conference on machine learning},
  \bibinfo{organization}{PMLR}. pp. \bibinfo{pages}{3929--3938}.
\bibitem[{He et~al.(2021)He, Neubig and Berg-Kirkpatrick}]{he2021efficient}
\bibinfo{author}{He, J.}, \bibinfo{author}{Neubig, G.},
  \bibinfo{author}{Berg-Kirkpatrick, T.}, \bibinfo{year}{2021}.
\newblock \bibinfo{title}{Efficient nearest neighbor language models}.
\newblock \bibinfo{journal}{arXiv preprint arXiv:2109.04212} .
\bibitem[{He et~al.(2023)He, Zhong, Cai, Lee and He}]{he2023rest}
\bibinfo{author}{He, Z.}, \bibinfo{author}{Zhong, Z.}, \bibinfo{author}{Cai,
  T.}, \bibinfo{author}{Lee, J.D.}, \bibinfo{author}{He, D.},
  \bibinfo{year}{2023}.
\newblock \bibinfo{title}{Rest: Retrieval-based speculative decoding}.
\newblock \bibinfo{journal}{arXiv preprint arXiv:2311.08252} .
\bibitem[{Hochreiter and Schmidhuber(1997)}]{hochreiter1997long}
\bibinfo{author}{Hochreiter, S.}, \bibinfo{author}{Schmidhuber, J.},
  \bibinfo{year}{1997}.
\newblock \bibinfo{title}{Long short-term memory}.
\newblock \bibinfo{journal}{Neural computation} \bibinfo{volume}{9},
  \bibinfo{pages}{1735--1780}.
\bibitem[{Houdt et~al.(2020)}]{lstm_survey}
\bibinfo{author}{Houdt, G.V.}, et~al., \bibinfo{year}{2020}.
\newblock \bibinfo{title}{A review on the long short-term memory model}.
\newblock \bibinfo{journal}{Artif. Intell. Rev.} \bibinfo{volume}{53},
  \bibinfo{pages}{5929--5955}.
\bibitem[{Hu et~al.(2021)Hu, Shen, Wallis, Allen-Zhu, Li, Wang, Wang and
  Chen}]{hu2021lora}
\bibinfo{author}{Hu, E.J.}, \bibinfo{author}{Shen, Y.},
  \bibinfo{author}{Wallis, P.}, \bibinfo{author}{Allen-Zhu, Z.},
  \bibinfo{author}{Li, Y.}, \bibinfo{author}{Wang, S.}, \bibinfo{author}{Wang,
  L.}, \bibinfo{author}{Chen, W.}, \bibinfo{year}{2021}.
\newblock \bibinfo{title}{Lora: Low-rank adaptation of large language models}.
\newblock \bibinfo{journal}{arXiv preprint arXiv:2106.09685} .
\bibitem[{Hu et~al.(2022)Hu, Wu, Shu and Qu}]{hu2022logical}
\bibinfo{author}{Hu, X.}, \bibinfo{author}{Wu, X.}, \bibinfo{author}{Shu, Y.},
  \bibinfo{author}{Qu, Y.}, \bibinfo{year}{2022}.
\newblock \bibinfo{title}{Logical form generation via multi-task learning for
  complex question answering over knowledge bases}, in:
  \bibinfo{booktitle}{Proceedings of the 29th International Conference on
  Computational Linguistics}, pp. \bibinfo{pages}{1687--1696}.
\bibitem[{Huang et~al.(2023)Huang, Huang, Yang, Ren, Liu, Li, Ye, Liu, Yin and
  Zhao}]{huang2023make}
\bibinfo{author}{Huang, R.}, \bibinfo{author}{Huang, J.},
  \bibinfo{author}{Yang, D.}, \bibinfo{author}{Ren, Y.}, \bibinfo{author}{Liu,
  L.}, \bibinfo{author}{Li, M.}, \bibinfo{author}{Ye, Z.},
  \bibinfo{author}{Liu, J.}, \bibinfo{author}{Yin, X.}, \bibinfo{author}{Zhao,
  Z.}, \bibinfo{year}{2023}.
\newblock \bibinfo{title}{Make-an-audio: Text-to-audio generation with
  prompt-enhanced diffusion models}, in: \bibinfo{booktitle}{International
  Conference on Machine Learning}, \bibinfo{organization}{PMLR}. pp.
  \bibinfo{pages}{13916--13932}.
\bibitem[{Huang et~al.(2021)Huang, Kim and Zou}]{DBLP:conf/emnlp/HuangKZ21}
\bibinfo{author}{Huang, X.}, \bibinfo{author}{Kim, J.}, \bibinfo{author}{Zou,
  B.}, \bibinfo{year}{2021}.
\newblock \bibinfo{title}{Unseen entity handling in complex question answering
  over knowledge base via language generation}, in: \bibinfo{booktitle}{EMNLP
  Findings}.
\bibitem[{Ilyas et~al.(2008)Ilyas, Beskales and Soliman}]{ilyas2008survey}
\bibinfo{author}{Ilyas, I.F.}, \bibinfo{author}{Beskales, G.},
  \bibinfo{author}{Soliman, M.A.}, \bibinfo{year}{2008}.
\newblock \bibinfo{title}{A survey of top-k query processing techniques in
  relational database systems}.
\newblock \bibinfo{journal}{ACM Computing Surveys (CSUR)} \bibinfo{volume}{40},
  \bibinfo{pages}{1--58}.
\bibitem[{Izacard and Grave(2020)}]{izacard2020leveraging}
\bibinfo{author}{Izacard, G.}, \bibinfo{author}{Grave, E.},
  \bibinfo{year}{2020}.
\newblock \bibinfo{title}{Leveraging passage retrieval with generative models
  for open domain question answering}.
\newblock \bibinfo{journal}{arXiv preprint arXiv:2007.01282} .
\bibitem[{Ji et~al.(2023)Ji, Lee, Frieske, Yu, Su, Xu, Ishii, Bang, Madotto and
  Fung}]{ji2023survey}
\bibinfo{author}{Ji, Z.}, \bibinfo{author}{Lee, N.}, \bibinfo{author}{Frieske,
  R.}, \bibinfo{author}{Yu, T.}, \bibinfo{author}{Su, D.}, \bibinfo{author}{Xu,
  Y.}, \bibinfo{author}{Ishii, E.}, \bibinfo{author}{Bang, Y.J.},
  \bibinfo{author}{Madotto, A.}, \bibinfo{author}{Fung, P.},
  \bibinfo{year}{2023}.
\newblock \bibinfo{title}{Survey of hallucination in natural language
  generation}.
\newblock \bibinfo{journal}{ACM Computing Surveys} \bibinfo{volume}{55},
  \bibinfo{pages}{1--38}.
\bibitem[{Jin et~al.(2024)Jin, Yang, Chen and Lu}]{jin2023genegpt}
\bibinfo{author}{Jin, Q.}, \bibinfo{author}{Yang, Y.}, \bibinfo{author}{Chen,
  Q.}, \bibinfo{author}{Lu, Z.}, \bibinfo{year}{2024}.
\newblock \bibinfo{title}{Genegpt: Augmenting large language models with domain
  tools for improved access to biomedical information}.
\newblock \bibinfo{journal}{Bioinformatics} \bibinfo{volume}{40},
  \bibinfo{pages}{btae075}.
\bibitem[{Jose et~al.(2024)Jose, Nguyen, Medjaher, Zemouri, L{\'e}vesque and
  Tahan}]{jose2024advancing}
\bibinfo{author}{Jose, S.}, \bibinfo{author}{Nguyen, K.T.},
  \bibinfo{author}{Medjaher, K.}, \bibinfo{author}{Zemouri, R.},
  \bibinfo{author}{L{\'e}vesque, M.}, \bibinfo{author}{Tahan, A.},
  \bibinfo{year}{2024}.
\newblock \bibinfo{title}{Advancing multimodal diagnostics: Integrating
  industrial textual data and domain knowledge with large language models}.
\newblock \bibinfo{journal}{Expert Systems with Applications} ,
  \bibinfo{pages}{124603}.
\bibitem[{Karras et~al.(2022)Karras, Aittala, Aila and
  Laine}]{karras2022elucidating}
\bibinfo{author}{Karras, T.}, \bibinfo{author}{Aittala, M.},
  \bibinfo{author}{Aila, T.}, \bibinfo{author}{Laine, S.},
  \bibinfo{year}{2022}.
\newblock \bibinfo{title}{Elucidating the design space of diffusion-based
  generative models}.
\newblock \bibinfo{journal}{Advances in neural information processing systems}
  \bibinfo{volume}{35}, \bibinfo{pages}{26565--26577}.
\bibitem[{Khandelwal et~al.(2019)Khandelwal, Levy, Jurafsky, Zettlemoyer and
  Lewis}]{khandelwal2019generalization}
\bibinfo{author}{Khandelwal, U.}, \bibinfo{author}{Levy, O.},
  \bibinfo{author}{Jurafsky, D.}, \bibinfo{author}{Zettlemoyer, L.},
  \bibinfo{author}{Lewis, M.}, \bibinfo{year}{2019}.
\newblock \bibinfo{title}{Generalization through memorization: Nearest neighbor
  language models}.
\newblock \bibinfo{journal}{arXiv preprint arXiv:1911.00172} .
\bibitem[{Kloeden et~al.(1992)Kloeden, Platen, Kloeden and
  Platen}]{kloeden1992stochastic}
\bibinfo{author}{Kloeden, P.E.}, \bibinfo{author}{Platen, E.},
  \bibinfo{author}{Kloeden, P.E.}, \bibinfo{author}{Platen, E.},
  \bibinfo{year}{1992}.
\newblock \bibinfo{title}{Stochastic differential equations}.
\newblock \bibinfo{publisher}{Springer}.
\bibitem[{Koizumi et~al.(2020)Koizumi, Ohishi, Niizumi, Takeuchi and
  Yasuda}]{koizumi2020audio}
\bibinfo{author}{Koizumi, Y.}, \bibinfo{author}{Ohishi, Y.},
  \bibinfo{author}{Niizumi, D.}, \bibinfo{author}{Takeuchi, D.},
  \bibinfo{author}{Yasuda, M.}, \bibinfo{year}{2020}.
\newblock \bibinfo{title}{Audio captioning using pre-trained large-scale
  language model guided by audio-based similar caption retrieval}.
\newblock \bibinfo{journal}{arXiv preprint arXiv:2012.07331} .
\bibitem[{Kucharavy(2024)}]{kucharavy2024deep}
\bibinfo{author}{Kucharavy, A.}, \bibinfo{year}{2024}.
\newblock \bibinfo{title}{From deep neural language models to {LLMs}}, in:
  \bibinfo{booktitle}{Large Language Models in Cybersecurity: Threats, Exposure
  and Mitigation}. \bibinfo{publisher}{Springer}, pp. \bibinfo{pages}{3--17}.
\bibitem[{Lafferty and Zhai(2001)}]{DBLP:conf/sigir/LaffertyZ01}
\bibinfo{author}{Lafferty, J.D.}, \bibinfo{author}{Zhai, C.},
  \bibinfo{year}{2001}.
\newblock \bibinfo{title}{Document language models, query models, and risk
  minimization for information retrieval}, in: \bibinfo{booktitle}{SIGIR}.
\bibitem[{Lei(2000)}]{lei2000industry}
\bibinfo{author}{Lei, D.T.}, \bibinfo{year}{2000}.
\newblock \bibinfo{title}{Industry evolution and competence development: the
  imperatives of technological convergence}.
\newblock \bibinfo{journal}{International Journal of Technology Management}
  \bibinfo{volume}{19}, \bibinfo{pages}{699--738}.
\bibitem[{Lewis et~al.(2020)Lewis, Perez, Piktus, Petroni, Karpukhin, Goyal,
  K{\"u}ttler, Lewis, Yih, Rockt{\"a}schel et~al.}]{lewis2020retrieval}
\bibinfo{author}{Lewis, P.}, \bibinfo{author}{Perez, E.},
  \bibinfo{author}{Piktus, A.}, \bibinfo{author}{Petroni, F.},
  \bibinfo{author}{Karpukhin, V.}, \bibinfo{author}{Goyal, N.},
  \bibinfo{author}{K{\"u}ttler, H.}, \bibinfo{author}{Lewis, M.},
  \bibinfo{author}{Yih, W.t.}, \bibinfo{author}{Rockt{\"a}schel, T.}, et~al.,
  \bibinfo{year}{2020}.
\newblock \bibinfo{title}{Retrieval-augmented generation for
  knowledge-intensive nlp tasks}.
\newblock \bibinfo{journal}{Advances in Neural Information Processing Systems}
  \bibinfo{volume}{33}, \bibinfo{pages}{9459--9474}.
\bibitem[{Li et~al.(2023)Li, Li, Savarese and Hoi}]{li2023blip}
\bibinfo{author}{Li, J.}, \bibinfo{author}{Li, D.}, \bibinfo{author}{Savarese,
  S.}, \bibinfo{author}{Hoi, S.}, \bibinfo{year}{2023}.
\newblock \bibinfo{title}{Blip-2: Bootstrapping language-image pre-training
  with frozen image encoders and large language models}, in:
  \bibinfo{booktitle}{International conference on machine learning},
  \bibinfo{organization}{PMLR}. pp. \bibinfo{pages}{19730--19742}.
\bibitem[{Li et~al.(2024)Li, Li, Zhang and Bian}]{li2024generalization}
\bibinfo{author}{Li, P.}, \bibinfo{author}{Li, Z.}, \bibinfo{author}{Zhang,
  H.}, \bibinfo{author}{Bian, J.}, \bibinfo{year}{2024}.
\newblock \bibinfo{title}{On the generalization properties of diffusion
  models}.
\newblock \bibinfo{journal}{Advances in Neural Information Processing Systems}
  \bibinfo{volume}{36}.
\bibitem[{Liang et~al.(2022)Liang, Zadeh and Morency}]{liang2022foundations}
\bibinfo{author}{Liang, P.P.}, \bibinfo{author}{Zadeh, A.},
  \bibinfo{author}{Morency, L.P.}, \bibinfo{year}{2022}.
\newblock \bibinfo{title}{Foundations and trends in multimodal machine
  learning: Principles, challenges, and open questions}.
\newblock \bibinfo{journal}{arXiv preprint arXiv:2209.03430} .
\bibitem[{Liu et~al.(2023a)Liu, Li, Li and Lee}]{liu2023improvedllava}
\bibinfo{author}{Liu, H.}, \bibinfo{author}{Li, C.}, \bibinfo{author}{Li, Y.},
  \bibinfo{author}{Lee, Y.J.}, \bibinfo{year}{2023}a.
\newblock \bibinfo{title}{Improved baselines with visual instruction tuning}.
\bibitem[{Liu et~al.(2024a)Liu, Li, Li, Li, Zhang, Shen and
  Lee}]{liu2024llavanext}
\bibinfo{author}{Liu, H.}, \bibinfo{author}{Li, C.}, \bibinfo{author}{Li, Y.},
  \bibinfo{author}{Li, B.}, \bibinfo{author}{Zhang, Y.}, \bibinfo{author}{Shen,
  S.}, \bibinfo{author}{Lee, Y.J.}, \bibinfo{year}{2024}a.
\newblock \bibinfo{title}{Llava-next: Improved reasoning, ocr, and world
  knowledge}.
\newblock \URLprefix
  \url{https://llava-vl.github.io/blog/2024-01-30-llava-next/}.
\bibitem[{Liu et~al.(2023b)Liu, Li, Wu and Lee}]{liu2023llava}
\bibinfo{author}{Liu, H.}, \bibinfo{author}{Li, C.}, \bibinfo{author}{Wu, Q.},
  \bibinfo{author}{Lee, Y.J.}, \bibinfo{year}{2023}b.
\newblock \bibinfo{title}{Visual instruction tuning}.
\bibitem[{Liu et~al.(2024b)Liu, Li, Wu and Lee}]{liu2024visual}
\bibinfo{author}{Liu, H.}, \bibinfo{author}{Li, C.}, \bibinfo{author}{Wu, Q.},
  \bibinfo{author}{Lee, Y.J.}, \bibinfo{year}{2024}b.
\newblock \bibinfo{title}{Visual instruction tuning}.
\newblock \bibinfo{journal}{Advances in neural information processing systems}
  \bibinfo{volume}{36}.
\bibitem[{Lu et~al.(2022)Lu, Mishra, Xia, Qiu, Chang, Zhu, Tafjord, Clark and
  Kalyan}]{lu2022learn}
\bibinfo{author}{Lu, P.}, \bibinfo{author}{Mishra, S.}, \bibinfo{author}{Xia,
  T.}, \bibinfo{author}{Qiu, L.}, \bibinfo{author}{Chang, K.W.},
  \bibinfo{author}{Zhu, S.C.}, \bibinfo{author}{Tafjord, O.},
  \bibinfo{author}{Clark, P.}, \bibinfo{author}{Kalyan, A.},
  \bibinfo{year}{2022}.
\newblock \bibinfo{title}{Learn to explain: Multimodal reasoning via thought
  chains for science question answering}.
\newblock \bibinfo{journal}{Advances in Neural Information Processing Systems}
  \bibinfo{volume}{35}, \bibinfo{pages}{2507--2521}.
\bibitem[{Löffler et~al.(2024)Löffler, Moskalenko, Herpich, Hanto, Hainsch,
  Bornemann, Diesing, Dupke and Barani}]{scenario}
\bibinfo{author}{Löffler, K.}, \bibinfo{author}{Moskalenko, N.},
  \bibinfo{author}{Herpich, P.}, \bibinfo{author}{Hanto, J.},
  \bibinfo{author}{Hainsch, K.}, \bibinfo{author}{Bornemann, J.},
  \bibinfo{author}{Diesing, A.}, \bibinfo{author}{Dupke, R.},
  \bibinfo{author}{Barani, M.}, \bibinfo{year}{2024}.
\newblock \bibinfo{title}{The European Energy Vision 2060 (EU EnVis-2060):
  Scenario Parametrization}.
\newblock \bibinfo{type}{Technical Report}. Man0EUvRE.
\newblock \DOIprefix\doi{10.5281/zenodo.13734444}. \bibinfo{note}{[Data set]}.
\bibitem[{Marcus et~al.(2022)Marcus, Davis and Aaronson}]{marcus2022very}
\bibinfo{author}{Marcus, G.}, \bibinfo{author}{Davis, E.},
  \bibinfo{author}{Aaronson, S.}, \bibinfo{year}{2022}.
\newblock \bibinfo{title}{A very preliminary analysis of dall-e 2}.
\newblock \bibinfo{journal}{arXiv preprint arXiv:2204.13807} .
\bibitem[{Masalkhi et~al.(2024)Masalkhi, Ong, Waisberg and
  Lee}]{masalkhi2024google}
\bibinfo{author}{Masalkhi, M.}, \bibinfo{author}{Ong, J.},
  \bibinfo{author}{Waisberg, E.}, \bibinfo{author}{Lee, A.G.},
  \bibinfo{year}{2024}.
\newblock \bibinfo{title}{Google deepmind’s gemini ai versus chatgpt: A
  comparative analysis in ophthalmology}.
\newblock \bibinfo{journal}{Eye} , \bibinfo{pages}{1--6}.
\bibitem[{Millinger et~al.(2022)Millinger, Tafarte, Jordan, Musonda, Chan,
  Meisel and Aliabadi}]{millinger2022model}
\bibinfo{author}{Millinger, M.}, \bibinfo{author}{Tafarte, P.},
  \bibinfo{author}{Jordan, M.}, \bibinfo{author}{Musonda, F.},
  \bibinfo{author}{Chan, K.}, \bibinfo{author}{Meisel, K.},
  \bibinfo{author}{Aliabadi, D.E.}, \bibinfo{year}{2022}.
\newblock \bibinfo{title}{A model for cost-and greenhouse gas optimal material
  and energy allocation of biomass and hydrogen}.
\newblock \bibinfo{journal}{SoftwareX} \bibinfo{volume}{20},
  \bibinfo{pages}{101264}.
\bibitem[{Najafi and Varol(2024)}]{najafi2024turkishbertweet}
\bibinfo{author}{Najafi, A.}, \bibinfo{author}{Varol, O.},
  \bibinfo{year}{2024}.
\newblock \bibinfo{title}{Turkishbertweet: Fast and reliable large language
  model for social media analysis}.
\newblock \bibinfo{journal}{Expert Systems with Applications} ,
  \bibinfo{pages}{124737}.
\bibitem[{Oh et~al.(2024)Oh, Park, Lim and Song}]{oh2024language}
\bibinfo{author}{Oh, C.}, \bibinfo{author}{Park, M.}, \bibinfo{author}{Lim,
  S.}, \bibinfo{author}{Song, K.}, \bibinfo{year}{2024}.
\newblock \bibinfo{title}{Language model-guided student performance prediction
  with multimodal auxiliary information}.
\newblock \bibinfo{journal}{Expert Systems with Applications}
  \bibinfo{volume}{250}, \bibinfo{pages}{123960}.
\bibitem[{Panda and Kaur(2023)}]{panda2023exploring}
\bibinfo{author}{Panda, S.}, \bibinfo{author}{Kaur, N.}, \bibinfo{year}{2023}.
\newblock \bibinfo{title}{Exploring the viability of chatgpt as an alternative
  to traditional chatbot systems in library and information centers}.
\newblock \bibinfo{journal}{Library hi tech news} \bibinfo{volume}{40},
  \bibinfo{pages}{22--25}.
\bibitem[{Parvez et~al.(2021)Parvez, Ahmad, Chakraborty, Ray and
  Chang}]{parvez2021retrieval}
\bibinfo{author}{Parvez, M.R.}, \bibinfo{author}{Ahmad, W.U.},
  \bibinfo{author}{Chakraborty, S.}, \bibinfo{author}{Ray, B.},
  \bibinfo{author}{Chang, K.W.}, \bibinfo{year}{2021}.
\newblock \bibinfo{title}{Retrieval augmented code generation and
  summarization}.
\newblock \bibinfo{journal}{arXiv preprint arXiv:2108.11601} .
\bibitem[{Proskurina et~al.(2024)Proskurina, Brun, Metzler and
  Velcin}]{proskurina2024quantization}
\bibinfo{author}{Proskurina, I.}, \bibinfo{author}{Brun, L.},
  \bibinfo{author}{Metzler, G.}, \bibinfo{author}{Velcin, J.},
  \bibinfo{year}{2024}.
\newblock \bibinfo{title}{When quantization affects confidence of large
  language models?}
\newblock \bibinfo{journal}{arXiv preprint arXiv:2405.00632} .
\bibitem[{Radford et~al.(2021)Radford, Kim, Hallacy, Ramesh, Goh, Agarwal,
  Sastry, Askell, Mishkin, Clark, Krueger and
  Sutskever}]{Radford2021LearningTV}
\bibinfo{author}{Radford, A.}, \bibinfo{author}{Kim, J.W.},
  \bibinfo{author}{Hallacy, C.}, \bibinfo{author}{Ramesh, A.},
  \bibinfo{author}{Goh, G.}, \bibinfo{author}{Agarwal, S.},
  \bibinfo{author}{Sastry, G.}, \bibinfo{author}{Askell, A.},
  \bibinfo{author}{Mishkin, P.}, \bibinfo{author}{Clark, J.},
  \bibinfo{author}{Krueger, G.}, \bibinfo{author}{Sutskever, I.},
  \bibinfo{year}{2021}.
\newblock \bibinfo{title}{Learning transferable visual models from natural
  language supervision}, in: \bibinfo{booktitle}{ICML}.
\bibitem[{Radford et~al.(2023)Radford, Kim, Xu, Brockman, McLeavey and
  Sutskever}]{radford2023robust}
\bibinfo{author}{Radford, A.}, \bibinfo{author}{Kim, J.W.},
  \bibinfo{author}{Xu, T.}, \bibinfo{author}{Brockman, G.},
  \bibinfo{author}{McLeavey, C.}, \bibinfo{author}{Sutskever, I.},
  \bibinfo{year}{2023}.
\newblock \bibinfo{title}{Robust speech recognition via large-scale weak
  supervision}, in: \bibinfo{booktitle}{International conference on machine
  learning}, \bibinfo{organization}{PMLR}. pp. \bibinfo{pages}{28492--28518}.
\bibitem[{Ramos et~al.(2023)Ramos, Martins, Elliott and
  Kementchedjhieva}]{ramos2023smallcap}
\bibinfo{author}{Ramos, R.}, \bibinfo{author}{Martins, B.},
  \bibinfo{author}{Elliott, D.}, \bibinfo{author}{Kementchedjhieva, Y.},
  \bibinfo{year}{2023}.
\newblock \bibinfo{title}{Smallcap: lightweight image captioning prompted with
  retrieval augmentation}, in: \bibinfo{booktitle}{Proceedings of the IEEE/CVF
  Conference on Computer Vision and Pattern Recognition}, pp.
  \bibinfo{pages}{2840--2849}.
\bibitem[{Robertson and Walker(1997)}]{DBLP:conf/sigir/RobertsonW97}
\bibinfo{author}{Robertson, S.E.}, \bibinfo{author}{Walker, S.},
  \bibinfo{year}{1997}.
\newblock \bibinfo{title}{On relevance weights with little relevance
  information}, in: \bibinfo{booktitle}{{SIGIR}}.
\bibitem[{Robertson and Zaragoza(2009)}]{DBLP:journals/ftir/RobertsonZ09}
\bibinfo{author}{Robertson, S.E.}, \bibinfo{author}{Zaragoza, H.},
  \bibinfo{year}{2009}.
\newblock \bibinfo{title}{The probabilistic relevance framework: {BM25} and
  beyond}.
\newblock \bibinfo{journal}{FTIR} \bibinfo{volume}{3},
  \bibinfo{pages}{333--389}.
\bibitem[{Rombach et~al.(2022a)Rombach, Blattmann, Lorenz, Esser and
  Ommer}]{rombach2022high}
\bibinfo{author}{Rombach, R.}, \bibinfo{author}{Blattmann, A.},
  \bibinfo{author}{Lorenz, D.}, \bibinfo{author}{Esser, P.},
  \bibinfo{author}{Ommer, B.}, \bibinfo{year}{2022}a.
\newblock \bibinfo{title}{High-resolution image synthesis with latent diffusion
  models}, in: \bibinfo{booktitle}{Proceedings of the IEEE/CVF conference on
  computer vision and pattern recognition}, pp. \bibinfo{pages}{10684--10695}.
\bibitem[{Rombach et~al.(2022b)Rombach, Blattmann, Lorenz
  et~al.}]{DBLP:conf/cvpr/RombachBLEO22}
\bibinfo{author}{Rombach, R.}, \bibinfo{author}{Blattmann, A.},
  \bibinfo{author}{Lorenz, D.}, et~al., \bibinfo{year}{2022}b.
\newblock \bibinfo{title}{High-resolution image synthesis with latent diffusion
  models}, in: \bibinfo{booktitle}{{IEEE/CVF}}.
\bibitem[{Rumelhart et~al.(1986)Rumelhart, Hinton and
  Williams}]{rumelhart1986learning}
\bibinfo{author}{Rumelhart, D.E.}, \bibinfo{author}{Hinton, G.E.},
  \bibinfo{author}{Williams, R.J.}, \bibinfo{year}{1986}.
\newblock \bibinfo{title}{Learning internal representations by error
  propagation, parallel distributed processing, explorations in the
  microstructure of cognition, ed. de rumelhart and j. mcclelland. vol. 1.
  1986}.
\newblock \bibinfo{journal}{Biometrika} \bibinfo{volume}{71},
  \bibinfo{pages}{6}.
\bibitem[{Sarto et~al.(2022)Sarto, Cornia, Baraldi and
  Cucchiara}]{sarto2022retrieval}
\bibinfo{author}{Sarto, S.}, \bibinfo{author}{Cornia, M.},
  \bibinfo{author}{Baraldi, L.}, \bibinfo{author}{Cucchiara, R.},
  \bibinfo{year}{2022}.
\newblock \bibinfo{title}{Retrieval-augmented transformer for image
  captioning}, in: \bibinfo{booktitle}{Proceedings of the 19th international
  conference on content-based multimedia indexing}, pp. \bibinfo{pages}{1--7}.
\bibitem[{Schuhmann et~al.(2022)Schuhmann, Beaumont, Vencu, Gordon, Wightman,
  Cherti, Coombes, Katta, Mullis, Wortsman, Schramowski, Kundurthy, Crowson,
  Schmidt, Kaczmarczyk and Jitsev}]{schuhmann2022laionb}
\bibinfo{author}{Schuhmann, C.}, \bibinfo{author}{Beaumont, R.},
  \bibinfo{author}{Vencu, R.}, \bibinfo{author}{Gordon, C.W.},
  \bibinfo{author}{Wightman, R.}, \bibinfo{author}{Cherti, M.},
  \bibinfo{author}{Coombes, T.}, \bibinfo{author}{Katta, A.},
  \bibinfo{author}{Mullis, C.}, \bibinfo{author}{Wortsman, M.},
  \bibinfo{author}{Schramowski, P.}, \bibinfo{author}{Kundurthy, S.R.},
  \bibinfo{author}{Crowson, K.}, \bibinfo{author}{Schmidt, L.},
  \bibinfo{author}{Kaczmarczyk, R.}, \bibinfo{author}{Jitsev, J.},
  \bibinfo{year}{2022}.
\newblock \bibinfo{title}{{LAION}-5b: An open large-scale dataset for training
  next generation image-text models}, in: \bibinfo{booktitle}{Thirty-sixth
  Conference on Neural Information Processing Systems Datasets and Benchmarks
  Track}.
\newblock \URLprefix \url{https://openreview.net/forum?id=M3Y74vmsMcY}.
\bibitem[{Seo et~al.(2024)Seo, Hong, Jang, Kim, Kwak, Lee and
  Kim}]{seo2024retrieval}
\bibinfo{author}{Seo, J.}, \bibinfo{author}{Hong, S.}, \bibinfo{author}{Jang,
  W.}, \bibinfo{author}{Kim, I.H.}, \bibinfo{author}{Kwak, M.},
  \bibinfo{author}{Lee, D.}, \bibinfo{author}{Kim, S.}, \bibinfo{year}{2024}.
\newblock \bibinfo{title}{Retrieval-augmented score distillation for text-to-3d
  generation}.
\newblock \bibinfo{journal}{arXiv preprint arXiv:2402.02972} .
\bibitem[{Tay et~al.(2023)Tay, Dehghani, Bahri and
  Metzler}]{EfficientTransformers}
\bibinfo{author}{Tay, Y.}, \bibinfo{author}{Dehghani, M.},
  \bibinfo{author}{Bahri, D.}, \bibinfo{author}{Metzler, D.},
  \bibinfo{year}{2023}.
\newblock \bibinfo{title}{Efficient transformers: {A} survey}.
\newblock \bibinfo{journal}{CSUR} \bibinfo{volume}{55},
  \bibinfo{pages}{109:1--109:28}.
\bibitem[{Thanh-Tung and Tran(2020)}]{thanh2020catastrophic}
\bibinfo{author}{Thanh-Tung, H.}, \bibinfo{author}{Tran, T.},
  \bibinfo{year}{2020}.
\newblock \bibinfo{title}{Catastrophic forgetting and mode collapse in {GANs}},
  in: \bibinfo{booktitle}{2020 international joint conference on neural
  networks (ijcnn)}, \bibinfo{organization}{IEEE}. pp. \bibinfo{pages}{1--10}.
\bibitem[{Touvron et~al.(2023)Touvron, Lavril, Izacard, Martinet, Lachaux,
  Lacroix, Rozi{\`e}re, Goyal, Hambro, Azhar et~al.}]{touvron2023llama}
\bibinfo{author}{Touvron, H.}, \bibinfo{author}{Lavril, T.},
  \bibinfo{author}{Izacard, G.}, \bibinfo{author}{Martinet, X.},
  \bibinfo{author}{Lachaux, M.A.}, \bibinfo{author}{Lacroix, T.},
  \bibinfo{author}{Rozi{\`e}re, B.}, \bibinfo{author}{Goyal, N.},
  \bibinfo{author}{Hambro, E.}, \bibinfo{author}{Azhar, F.}, et~al.,
  \bibinfo{year}{2023}.
\newblock \bibinfo{title}{Llama: Open and efficient foundation language
  models}.
\newblock \bibinfo{journal}{arXiv preprint arXiv:2302.13971} .
\bibitem[{Tseng et~al.(2020)Tseng, Lee, Jiang, Yang and
  Yang}]{tseng2020retrievegan}
\bibinfo{author}{Tseng, H.Y.}, \bibinfo{author}{Lee, H.Y.},
  \bibinfo{author}{Jiang, L.}, \bibinfo{author}{Yang, M.H.},
  \bibinfo{author}{Yang, W.}, \bibinfo{year}{2020}.
\newblock \bibinfo{title}{Retrievegan: Image synthesis via differentiable patch
  retrieval}, in: \bibinfo{booktitle}{Computer Vision--ECCV 2020: 16th European
  Conference, Glasgow, UK, August 23--28, 2020, Proceedings, Part VIII 16},
  \bibinfo{organization}{Springer}. pp. \bibinfo{pages}{242--257}.
\bibitem[{Urban et~al.(2024)Urban, D{\v{e}}cht{\v{e}}renko, Lukavsk{\`y},
  Hrabalov{\'a}, Svacha, Brom and Urban}]{urban2024chatgpt}
\bibinfo{author}{Urban, M.}, \bibinfo{author}{D{\v{e}}cht{\v{e}}renko, F.},
  \bibinfo{author}{Lukavsk{\`y}, J.}, \bibinfo{author}{Hrabalov{\'a}, V.},
  \bibinfo{author}{Svacha, F.}, \bibinfo{author}{Brom, C.},
  \bibinfo{author}{Urban, K.}, \bibinfo{year}{2024}.
\newblock \bibinfo{title}{Chatgpt improves creative problem-solving performance
  in university students: An experimental study}.
\newblock \bibinfo{journal}{Computers \& Education} \bibinfo{volume}{215},
  \bibinfo{pages}{105031}.
\bibitem[{Vaswani et~al.(2017)Vaswani, Shazeer, Parmar, Uszkoreit, Jones,
  Gomez, Kaiser and Polosukhin}]{vaswani2017attention}
\bibinfo{author}{Vaswani, A.}, \bibinfo{author}{Shazeer, N.},
  \bibinfo{author}{Parmar, N.}, \bibinfo{author}{Uszkoreit, J.},
  \bibinfo{author}{Jones, L.}, \bibinfo{author}{Gomez, A.N.},
  \bibinfo{author}{Kaiser, {\L}.}, \bibinfo{author}{Polosukhin, I.},
  \bibinfo{year}{2017}.
\newblock \bibinfo{title}{Attention is all you need}.
\newblock \bibinfo{journal}{Advances in neural information processing systems}
  \bibinfo{volume}{30}.
\bibitem[{Wang et~al.(2024)Wang, Ong, Wang, Ong, Cheng and
  Ong}]{wang2024potential}
\bibinfo{author}{Wang, C.}, \bibinfo{author}{Ong, J.}, \bibinfo{author}{Wang,
  C.}, \bibinfo{author}{Ong, H.}, \bibinfo{author}{Cheng, R.},
  \bibinfo{author}{Ong, D.}, \bibinfo{year}{2024}.
\newblock \bibinfo{title}{Potential for gpt technology to optimize future
  clinical decision-making using retrieval-augmented generation}.
\newblock \bibinfo{journal}{Annals of Biomedical Engineering}
  \bibinfo{volume}{52}, \bibinfo{pages}{1115--1118}.
\bibitem[{Wang et~al.(2022)Wang, Nie, Qiao et~al.}]{wang2022retrieval}
\bibinfo{author}{Wang, Z.}, \bibinfo{author}{Nie, W.}, \bibinfo{author}{Qiao,
  Z.}, et~al., \bibinfo{year}{2022}.
\newblock \bibinfo{title}{Retrieval-based controllable molecule generation},
  in: \bibinfo{booktitle}{ICLR}.
\bibitem[{Wu and Chen(2020)}]{wu2020systematic}
\bibinfo{author}{Wu, J.}, \bibinfo{author}{Chen, D.T.V.}, \bibinfo{year}{2020}.
\newblock \bibinfo{title}{A systematic review of educational digital
  storytelling}.
\newblock \bibinfo{journal}{Computers \& Education} \bibinfo{volume}{147},
  \bibinfo{pages}{103786}.
\bibitem[{Wu et~al.(2023a)Wu, Gan, Chen, Wan and Lin}]{wu2023ai}
\bibinfo{author}{Wu, J.}, \bibinfo{author}{Gan, W.}, \bibinfo{author}{Chen,
  Z.}, \bibinfo{author}{Wan, S.}, \bibinfo{author}{Lin, H.},
  \bibinfo{year}{2023}a.
\newblock \bibinfo{title}{Ai-generated content (aigc): A survey}.
\newblock \bibinfo{journal}{arXiv preprint arXiv:2304.06632} .
\bibitem[{Wu et~al.(2023b)Wu, Gan, Chen, Wan and Philip}]{wu2023multimodal}
\bibinfo{author}{Wu, J.}, \bibinfo{author}{Gan, W.}, \bibinfo{author}{Chen,
  Z.}, \bibinfo{author}{Wan, S.}, \bibinfo{author}{Philip, S.Y.},
  \bibinfo{year}{2023}b.
\newblock \bibinfo{title}{Multimodal large language models: A survey}, in:
  \bibinfo{booktitle}{2023 IEEE International Conference on Big Data
  (BigData)}, \bibinfo{organization}{IEEE}. pp. \bibinfo{pages}{2247--2256}.
\bibitem[{Xia et~al.(2015)Xia, Xu, Lan, Guo and Cheng}]{xia2015learning}
\bibinfo{author}{Xia, L.}, \bibinfo{author}{Xu, J.}, \bibinfo{author}{Lan, Y.},
  \bibinfo{author}{Guo, J.}, \bibinfo{author}{Cheng, X.}, \bibinfo{year}{2015}.
\newblock \bibinfo{title}{Learning maximal marginal relevance model via
  directly optimizing diversity evaluation measures}, in:
  \bibinfo{booktitle}{Proceedings of the 38th international ACM SIGIR
  conference on research and development in information retrieval}, pp.
  \bibinfo{pages}{113--122}.
\bibitem[{Xing et~al.(2024)Xing, Shi, Huang, Wu, Nan, Zhang, Fang, Roberts,
  Sch{\"o}nlieb, Del~Ser et~al.}]{xing2024ai}
\bibinfo{author}{Xing, X.}, \bibinfo{author}{Shi, F.}, \bibinfo{author}{Huang,
  J.}, \bibinfo{author}{Wu, Y.}, \bibinfo{author}{Nan, Y.},
  \bibinfo{author}{Zhang, S.}, \bibinfo{author}{Fang, Y.},
  \bibinfo{author}{Roberts, M.}, \bibinfo{author}{Sch{\"o}nlieb, C.B.},
  \bibinfo{author}{Del~Ser, J.}, et~al., \bibinfo{year}{2024}.
\newblock \bibinfo{title}{When ai eats itself: On the caveats of data pollution
  in the era of generative ai}.
\newblock \bibinfo{journal}{arXiv preprint arXiv:2405.09597} .
\bibitem[{Xu et~al.(2024a)Xu, Huang, Hou, Chen, Zhang, Feng and
  Xie}]{xu2024retrieval}
\bibinfo{author}{Xu, J.}, \bibinfo{author}{Huang, Y.}, \bibinfo{author}{Hou,
  J.}, \bibinfo{author}{Chen, G.}, \bibinfo{author}{Zhang, Y.},
  \bibinfo{author}{Feng, R.}, \bibinfo{author}{Xie, W.}, \bibinfo{year}{2024}a.
\newblock \bibinfo{title}{Retrieval-augmented egocentric video captioning}, in:
  \bibinfo{booktitle}{Proceedings of the IEEE/CVF Conference on Computer Vision
  and Pattern Recognition}, pp. \bibinfo{pages}{13525--13536}.
\bibitem[{Xu et~al.(2024b)Xu, Chen and Chen}]{xu2024enhancing}
\bibinfo{author}{Xu, S.}, \bibinfo{author}{Chen, M.}, \bibinfo{author}{Chen,
  S.}, \bibinfo{year}{2024}b.
\newblock \bibinfo{title}{Enhancing retrieval-augmented generation models with
  knowledge graphs: Innovative practices through a dual-pathway approach}, in:
  \bibinfo{booktitle}{International Conference on Intelligent Computing},
  \bibinfo{organization}{Springer}. pp. \bibinfo{pages}{398--409}.
\bibitem[{Yan et~al.(2021)Yan, Zhang, Abbeel and Srinivas}]{yan2021videogpt}
\bibinfo{author}{Yan, W.}, \bibinfo{author}{Zhang, Y.},
  \bibinfo{author}{Abbeel, P.}, \bibinfo{author}{Srinivas, A.},
  \bibinfo{year}{2021}.
\newblock \bibinfo{title}{Videogpt: Video generation using vq-vae and
  transformers}.
\newblock \href{http://arxiv.org/abs/2104.10157}{\tt arXiv:2104.10157}.
\bibitem[{Yang et~al.(2023a)Yang, Zhang, Song, Hong, Xu, Zhao, Zhang, Cui and
  Yang}]{yang2023diffusion}
\bibinfo{author}{Yang, L.}, \bibinfo{author}{Zhang, Z.}, \bibinfo{author}{Song,
  Y.}, \bibinfo{author}{Hong, S.}, \bibinfo{author}{Xu, R.},
  \bibinfo{author}{Zhao, Y.}, \bibinfo{author}{Zhang, W.},
  \bibinfo{author}{Cui, B.}, \bibinfo{author}{Yang, M.H.},
  \bibinfo{year}{2023}a.
\newblock \bibinfo{title}{Diffusion models: A comprehensive survey of methods
  and applications}.
\newblock \bibinfo{journal}{ACM Computing Surveys} \bibinfo{volume}{56},
  \bibinfo{pages}{1--39}.
\bibitem[{Yang et~al.(2023b)Yang, Zhang et~al.}]{yang2023diffsurvey}
\bibinfo{author}{Yang, L.}, \bibinfo{author}{Zhang, Z.}, et~al.,
  \bibinfo{year}{2023}b.
\newblock \bibinfo{title}{Diffusion models: A comprehensive survey of methods
  and applications}.
\newblock \bibinfo{journal}{CSUR} \bibinfo{volume}{56}, \bibinfo{pages}{1--39}.
\bibitem[{Ye et~al.(2023)Ye, Xu, Xu, Ye, Yan, Zhou, Wang, Hu, Shi, Shi
  et~al.}]{ye2023mplug}
\bibinfo{author}{Ye, Q.}, \bibinfo{author}{Xu, H.}, \bibinfo{author}{Xu, G.},
  \bibinfo{author}{Ye, J.}, \bibinfo{author}{Yan, M.}, \bibinfo{author}{Zhou,
  Y.}, \bibinfo{author}{Wang, J.}, \bibinfo{author}{Hu, A.},
  \bibinfo{author}{Shi, P.}, \bibinfo{author}{Shi, Y.}, et~al.,
  \bibinfo{year}{2023}.
\newblock \bibinfo{title}{mplug-owl: Modularization empowers large language
  models with multimodality}.
\newblock \bibinfo{journal}{arXiv preprint arXiv:2304.14178} .
\bibitem[{Yu et~al.(2024)Yu, Gan, Zhang, Tong, Liu and Liu}]{yu2024evaluation}
\bibinfo{author}{Yu, H.}, \bibinfo{author}{Gan, A.}, \bibinfo{author}{Zhang,
  K.}, \bibinfo{author}{Tong, S.}, \bibinfo{author}{Liu, Q.},
  \bibinfo{author}{Liu, Z.}, \bibinfo{year}{2024}.
\newblock \bibinfo{title}{Evaluation of retrieval-augmented generation: A
  survey}.
\newblock \bibinfo{journal}{arXiv preprint arXiv:2405.07437} .
\bibitem[{Zamani et~al.(2022)Zamani, Diaz, Dehghani, Metzler and
  Bendersky}]{zamani2022retrieval}
\bibinfo{author}{Zamani, H.}, \bibinfo{author}{Diaz, F.},
  \bibinfo{author}{Dehghani, M.}, \bibinfo{author}{Metzler, D.},
  \bibinfo{author}{Bendersky, M.}, \bibinfo{year}{2022}.
\newblock \bibinfo{title}{Retrieval-enhanced machine learning}, in:
  \bibinfo{booktitle}{Proceedings of the 45th International ACM SIGIR
  Conference on Research and Development in Information Retrieval}, pp.
  \bibinfo{pages}{2875--2886}.
\bibitem[{Zhang et~al.(2023a)Zhang, Rao and Agrawala}]{zhang2023adding}
\bibinfo{author}{Zhang, L.}, \bibinfo{author}{Rao, A.},
  \bibinfo{author}{Agrawala, M.}, \bibinfo{year}{2023}a.
\newblock \bibinfo{title}{Adding conditional control to text-to-image diffusion
  models}, in: \bibinfo{booktitle}{Proceedings of the IEEE/CVF International
  Conference on Computer Vision}, pp. \bibinfo{pages}{3836--3847}.
\bibitem[{Zhang et~al.(2023b)Zhang, Guo, Pan, Cai, Hong, Li, Yang and
  Liu}]{zhang2023remodiffuse}
\bibinfo{author}{Zhang, M.}, \bibinfo{author}{Guo, X.}, \bibinfo{author}{Pan,
  L.}, \bibinfo{author}{Cai, Z.}, \bibinfo{author}{Hong, F.},
  \bibinfo{author}{Li, H.}, \bibinfo{author}{Yang, L.}, \bibinfo{author}{Liu,
  Z.}, \bibinfo{year}{2023}b.
\newblock \bibinfo{title}{Remodiffuse: Retrieval-augmented motion diffusion
  model}, in: \bibinfo{booktitle}{Proceedings of the IEEE/CVF International
  Conference on Computer Vision}, pp. \bibinfo{pages}{364--373}.
\bibitem[{Zhao et~al.(2024)Zhao, Zhang, Yu, Wang, Geng, Fu, Yang, Zhang and
  Cui}]{zhao2024retrieval}
\bibinfo{author}{Zhao, P.}, \bibinfo{author}{Zhang, H.}, \bibinfo{author}{Yu,
  Q.}, \bibinfo{author}{Wang, Z.}, \bibinfo{author}{Geng, Y.},
  \bibinfo{author}{Fu, F.}, \bibinfo{author}{Yang, L.}, \bibinfo{author}{Zhang,
  W.}, \bibinfo{author}{Cui, B.}, \bibinfo{year}{2024}.
\newblock \bibinfo{title}{Retrieval-augmented generation for ai-generated
  content: A survey}.
\newblock \bibinfo{journal}{arXiv preprint arXiv:2402.19473} .
\bibitem[{Zhou et~al.(2022)Zhou, Alon, Xu, Wang, Jiang and
  Neubig}]{zhou2022docprompting}
\bibinfo{author}{Zhou, S.}, \bibinfo{author}{Alon, U.}, \bibinfo{author}{Xu,
  F.F.}, \bibinfo{author}{Wang, Z.}, \bibinfo{author}{Jiang, Z.},
  \bibinfo{author}{Neubig, G.}, \bibinfo{year}{2022}.
\newblock \bibinfo{title}{Docprompting: Generating code by retrieving the
  docs}.
\newblock \bibinfo{journal}{arXiv preprint arXiv:2207.05987} .
\bibitem[{Zhu et~al.(2023)Zhu, Chen, Shen, Li and Elhoseiny}]{zhu2023minigpt}
\bibinfo{author}{Zhu, D.}, \bibinfo{author}{Chen, J.}, \bibinfo{author}{Shen,
  X.}, \bibinfo{author}{Li, X.}, \bibinfo{author}{Elhoseiny, M.},
  \bibinfo{year}{2023}.
\newblock \bibinfo{title}{Minigpt-4: Enhancing vision-language understanding
  with advanced large language models}.
\newblock \bibinfo{journal}{arXiv preprint arXiv:2304.10592} .

\end{thebibliography}


\bio{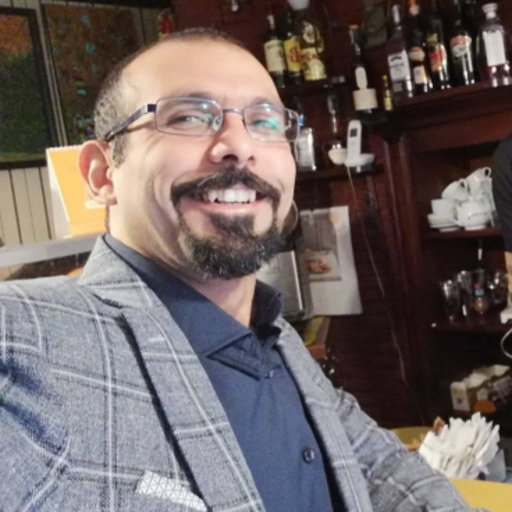}
Ali Forootani received the M.Sc. degree in electrical engineering and automatic control systems from the Power and Water University of Technology (Iran) in 2011, and the Ph.D. degree from the Department of Engineering, University of Sannio (Italy) in 2019. From 2011 to 2015 he worked both on research and industry at Niroo Research Institute (Iran) and at the Ministry of Energy and Power (Iran). From 2019 to 2024, he served as a Postdoctoral researcher at the University of Sannio, the University of Salerno, at Maynooth University (Ireland), and the Max Planck Institute for Dynamics of Complex Technical Systems in Magdeburg (Germany). Since May 2024, he has been affiliated with the Helmholtz Center for Environmental Research-UFZ in Leipzig (Germany). His current research interests include Markov decision processes, approximate dynamic programming, reinforcement learning in optimal control, network control systems, Physics Informed Neural Networks (PINNs), PDE parameter estimation and identification, and Large Language Models. He is a Senior Member of IEEE Control System Society.
\endbio

\bio{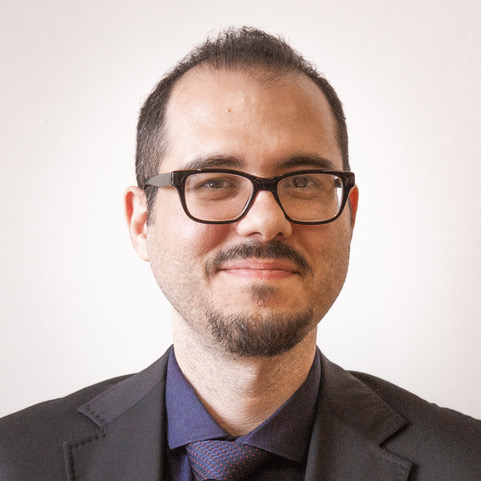}
Danial Esmaeili Aliabadi leads the ``Renewable Energies" working group at the Helmholtz Centre for Environmental Research (UFZ), the Department of Bioenergy. His research primarily revolves around bioenergy, with a broader interest in energy markets, energy policy, and sustainable development. Prior to his position at UFZ, from 2017 to 2020, he was the lead researcher at the Istanbul International Center for Energy and Climate (IICEC), developing the IICEC-Sabanci University TIMES Energy Model (ISTEM). Dr. Esmaeili completed his Ph.D. at Sabanci University, Istanbul, focusing on electricity markets and energy planning. He also obtained his M.Sc. in Industrial Engineering from Qazvin Azad University, Iran, with many national and international awards in computer sciences and robotics.
\endbio

\bio{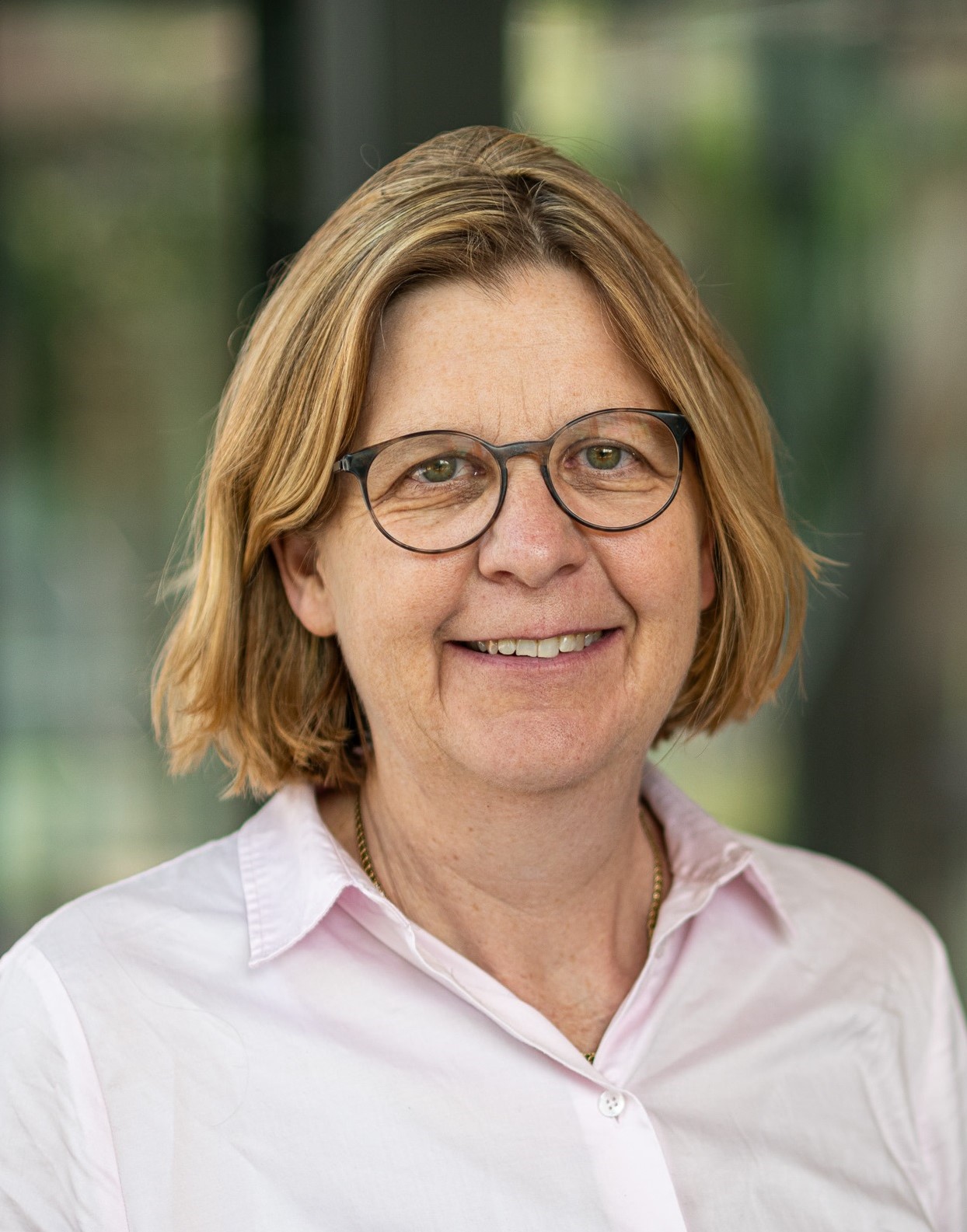}
Daniela Thr\"an studied technical environmental protection at the University of Berlin and received her doctorate at the Bauhaus University Weimar. Since 2009 she has led different teams at the Deutsche Biomasseforschungszentrum and at the Helmholtz Center for Environmental Research (UFZ) in Leipzig in the areas of biomass potential, energetic and material use of biomass, material flow analyses, and sustainability issues. Since the end of 2011, Daniela Thrän has held the chair in bioenergy systems at the Faculty of Economics at the University of Leipzig. She is task leader of IEA Bioenergy Task 44 Flexible Bioenergy and System Integration and she has been the department head of ``Bioenergy Systems" at DBFZ – Deutsches Biomasseforschungszentrum gGmbH from 2003 until 2023. Currently, she heads the Department of Bioenergy at the Helmholtz Centre for Environmental Research (UFZ).
\endbio

\end{document}